\documentclass[sigconf, nonacm]{acmart}

\usepackage{subfig}
\usepackage{algorithm}
\usepackage{algpseudocode} 
\usepackage{listings}
\usepackage{hyperref}
\usepackage{enumitem}
\usepackage{tabularx}
\usepackage{array}
\makeatletter
\g@addto@macro{\endtabular}{\rowfont{}}
\makeatother
\newcommand{\rowfonttype}{}
\newcommand{\rowfont}[1]{
   \gdef\rowfonttype{#1}#1%
}
\newcolumntype{Y}{>{\centering\arraybackslash\rowfonttype}X}

\newcommand{\tp}{\emph{(topic, partition)}}

\newcommand\vldbdoi{XX.XX/XXX.XX}
\newcommand\vldbpages{XXX-XXX}
\newcommand\vldbvolume{14}
\newcommand\vldbissue{1}
\newcommand\vldbyear{2020}
\newcommand\vldbauthors{\authors}
\newcommand\vldbtitle{\shorttitle} 
\newcommand\vldbavailabilityurl{http://vldb.org/pvldb/format_vol14.html}
\newcommand\vldbpagestyle{plain} 

\begin{document}
\title{Railgun: managing large streaming windows
under MAD requirements}
\author{Ana Sofia Gomes}
\email{sofia.gomes@feedzai.com}
\affiliation{%
  \institution{Feedzai}
}

\author{João Oliveirinha}
\email{joao.oliveirinha@feedzai.com}
\affiliation{%
  \institution{Feedzai}
}

\author{Pedro Cardoso}
\email{pedro.cardoso@feedzai.com}
\affiliation{%
  \institution{Feedzai}
}

\author{Pedro Bizarro}
\email{pedro.bizarro@feedzai.com}
\affiliation{%
  \institution{Feedzai}
}

\begin{abstract}
Some mission critical systems, e.g., fraud detection, require accurate, real-time metrics over long time sliding windows on applications that demand high throughput and low latencies. 
As these applications need to run ``forever'' and cope with large, spiky data loads, they further require to be run in a distributed setting.
We are unaware of any streaming system that provides all those properties. Instead, existing systems take large simplifications, such as implementing sliding windows as a fixed set of overlapping windows, jeopardizing metric accuracy (violating regulatory rules) or latency (breaching service agreements).
In this paper, we propose Railgun, a fault-tolerant, elastic, and distributed streaming system supporting real-time sliding windows for scenarios requiring high loads and millisecond-level latencies. We benchmarked an initial prototype of Railgun using real data, showing significant lower latency than Flink and low memory usage independent of window size. Further, we show that Railgun scales nearly linearly, respecting our msec-level latencies at high percentiles (<250ms @ 99.9\%) even under a load of 1 million events per second.
\end{abstract}

\maketitle

\pagestyle{\vldbpagestyle}
\begingroup\small\noindent\raggedright\textbf{PVLDB Reference Format:}\\
\vldbauthors. \vldbtitle. PVLDB, \vldbvolume(\vldbissue): \vldbpages, \vldbyear.\\
\href{https://doi.org/\vldbdoi}{doi:\vldbdoi}
\endgroup
\begingroup
\renewcommand\thefootnote{}\footnote{\noindent
This work is licensed under the Creative Commons BY-NC-ND 4.0 International License. Visit \url{https://creativecommons.org/licenses/by-nc-nd/4.0/} to view a copy of this license. For any use beyond those covered by this license, obtain permission by emailing \href{mailto:info@vldb.org}{info@vldb.org}. Copyright is held by the owner/author(s). Publication rights licensed to the VLDB Endowment. \\
\raggedright Proceedings of the VLDB Endowment, Vol. \vldbvolume, No. \vldbissue\ %
ISSN 2150-8097. \\
\href{https://doi.org/\vldbdoi}{doi:\vldbdoi} \\
}\addtocounter{footnote}{-1}\endgroup

\ifdefempty{\vldbavailabilityurl}{}{
\vspace{.3cm}
\begingroup\small\noindent\raggedright\textbf{PVLDB Artifact Availability:}\\
The source code, data, and/or other artifacts have been made available at \url{\vldbavailabilityurl}.
\endgroup
}

\section{Introduction}
\label{sec:introduction}

In some mission critical systems, e.g., financial fraud detection, it is desirable that the underlying streaming engines fulfill our proposed M-A-D requirements:
\begin{itemize}
  \item[] \textbf{M}sec-level latencies at high percentiles (<250ms @ 99.9\%);
  \item[] \textbf{A}ccurate sliding window aggregations event-by-event;
  \item[] \textbf{D}istributed, scalable and fault-tolerant.
\end{itemize}

However, to the best of our knowledge, no streaming engine today delivers all three MAD requirements. Initial streaming engines such as STREAM~\cite{DBLP:conf/sigmod/ArasuBBDIRW03}, NiagaraCQ~\cite{niagaracq} or Siddhi~\cite{suhothayan2011siddhi} provide accurate sliding window aggregations per event, but do not scale beyond one node, nor do they comply with millisecond-level latencies.
In opposition, state-of-the-art streaming engines such as Flink~\cite{DBLP:journals/debu/CarboneKEMHT15}, Kafka Streams~\cite{kreps2016introducing}, and others \cite{sparkstreaming, Akka,noghabi2017samza,ramasamy2019unifying} provide scalability and fault-tolerance with low latencies, but at the expense of inaccurate sliding windows aggregations due to their window choices or load shedding~\cite{DBLP:conf/cidr/AbadiABCCHLMRRTXZ05}, failing to meet \textbf{A}.

A critical decision is how to handle a \emph{large streaming state} while delivering \emph{low latency}. In low throughput and small windows, events can fit in-memory of a single node, and accurate aggregations can be computed for every new event over sliding windows. However, for large windows or high throughput (where \textbf{D} is required) handling the incoming \emph{and} expiring events becomes such a problem that streaming engines either shed load, or use hopping windows as an \emph{approximation} of real-time sliding windows, computing aggregations and expiring events only so often.

\begin{figure}[hb!]
    \centering
    \includegraphics[scale=0.7]{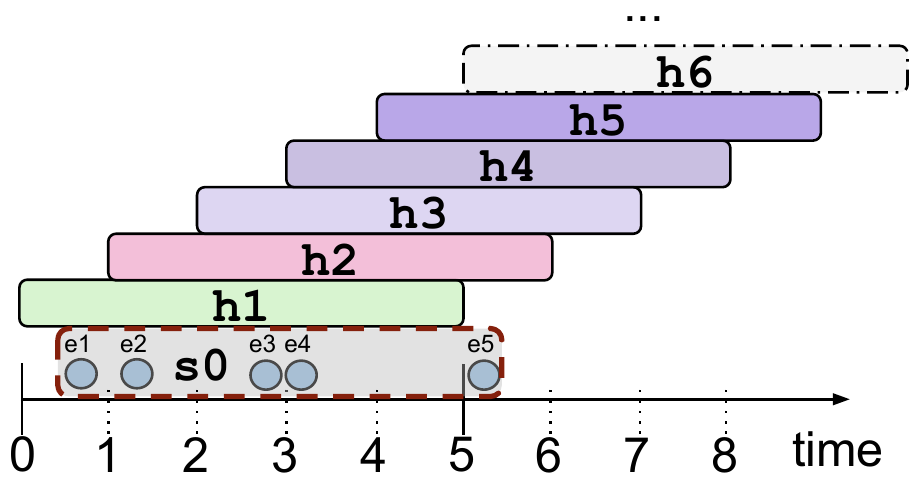}
    \caption{A 5-min hopping window with 1-min hop uses physical windows (h1-h6) but none capture the 5 events (circles) together, unlike a real-time sliding window (s0).}
    \label{fig:hop-example}
\end{figure}

For instance, using hopping windows, a 5-min sliding window can be approximated, e.g., using five fixed physical 5-min windows, each offset by 1 minute (the \emph{hop}) and where, as time passes, new windows (and their aggregations) are created and expired. Figure~\ref{fig:hop-example} illustrates this behavior of hopping windows, and how they might lead to inaccurate aggregations. At timestamp 5, there are exactly 5 active physical windows, h1-h5. When $e_5$ arrives, still within a 5-min window of $e_1$, window h1 is already expired and h6 created. Since h1 expires at timestamp 5, and h2 only starts at timestamp 2, both h1 and h2 only count 4 events, e1-e4 and e2-e5, respectively. 
This shows an example where a true 5-min sliding window (s0) takes into account all 5 events for its aggregations when $e_5$ arrives, but a hopping window with 1-min hop does not. The hop could be made smaller, e.g., 1 second, but that would imply concurrently managing 300 5-min physical windows, instead of 5. Additionally, since the frequency on which time slides is still fixed, a 1 second hop window might still not capture all 5 events together.

\begin{figure}[b]
    \centering
      \includegraphics[width=0.45\textwidth]{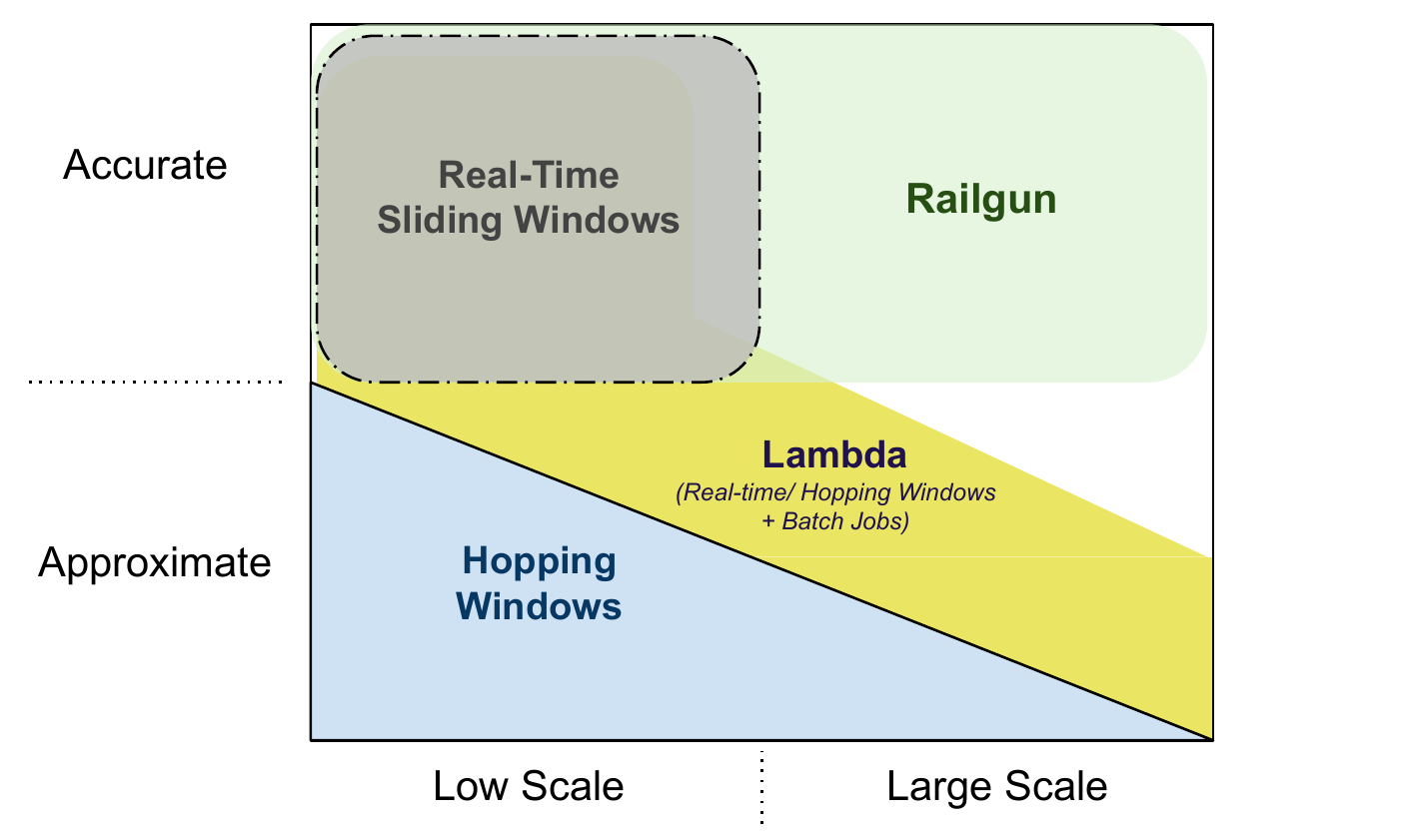}
    \caption{Approaches to manage \emph{very large} streaming state to fulfill tail low latency.}
    \label{fig:approach-table}
\end{figure}

To keep latency low while dealing with high event throughput, state-of-the-art streaming engines use hopping windows in an attempt to save memory. Aggregations over real-time sliding windows require accessing all events to compute accurate aggregations, making these solutions low scale. Approximate aggregations can be done over hopping windows by discarding event tuples, but require handling the multiple aggregation window states, where the number of window states is defined by a ratio between the window size, and the hop size.
As we shall see in Section~\ref{subsection:flinkVsRailgun}, this tradeoff works until the hop size is much smaller than the window size. When windows start to span over multiple minutes, or hours, the window aggregation state becomes so large, that streaming engines either choose to reduce aggregation precision further (by using larger hops of minutes or hours), or deploy lambda architectures, to combine delayed results computed in batch with small real-time windows (see Figure~\ref{fig:approach-table}).

This state of affairs presents a challenge for modern fraud detection systems. Responsible for processing trillions of dollars per year worldwide, these mission-critical systems have demanding latency requirements (e.g., <250 ms for  99.9\% percentile), and still require accurate aggregation metrics event-by-event (for regulatory and adversarial reasons, see Section~\ref{sec:realtimewindows}) over long windows.
To address this need, we propose Railgun, a novel distributed streaming engine based on low-memory-footprint, disk-backed sliding windows (an improvement on the sliding windows and SlideM algorithm~\cite{slidem}) on top of which, we built state-management and distributed communication layers to fulfill all of the MAD requirements. Our contributions are as follows:
\begin{enumerate}
    \item We formulate the MAD requirements, supporting why they are needed in use-cases such as fraud-detection (Section~\ref{sec:realtimewindows});
    \item We present our proposal, Railgun, with an overview of the architecture, components and decisions (Sections~\ref{sec:railgun} and \ref{sec:computation});
    \item We illustrate how Flink degrades when small hops are used to approximate real-time sliding windows (Section~\ref{subsection:flinkVsRailgun});
    \item We show that Railgun computes real-time metrics over large windows in an efficient and scalable way (Section~\ref{subsec:scalingRailgun}).
    \item We show that Railgun scales nearly linearly, up to 50 nodes, even under a load of 1 million events per second, while respecting our latency requirement at high percentiles (Section~\ref{subsec:distributedRailgun}).
\end{enumerate}

\section{Background}
\label{sec:background}
A data stream $S$ is an unbounded sequence of events $e_1, e_2, ...$, each with a timestamp. Aggregations over streams are computed using windows. A \emph{window} $w$ is a sequence of contiguous events of $S$ with a certain size $w_s$ (defined by a number of events, time interval, or start-stop conditions as in a user session). In this paper, we focus on time-based windows, henceforth referred simply as windows.
As time passes, a window over a stream is evaluated often, at a specific, and changing, timepoint $T_{eval}$. $T_{eval}$ determines the events to include for the aggregations, where an event with timestamp $t_i$ belongs to a window evaluation iff $T_{eval} - w_s \leq t_i < T_{eval}$.

\emph{Hopping windows} are windows where $T_{eval}$ changes according to a step of length $s$.
This step $s$, or \emph{hop}, marks \emph{when} new windows are created. If $s$ is smaller than $w_s$, then the windows overlap, i.e., an event may belong to more than one hopping window\footnote{Hopping windows are often called sliding windows by systems such as Flink because they \emph{approximate} the behavior of real-time sliding windows.}. When $s$ is equal to $w_s$, hopping windows do not overlap, and events belong to exactly one window. This case is frequently given the name of \emph{tumbling windows}. Step $s$ is generally not bigger than $w_s$.

\emph{Real-time sliding windows}, or just \emph{sliding windows}, are windows where $T_{eval}$ is the moment right after a new event has arrived. This frequent evaluation is computationally expensive as, for each new event $e_i$, the system has to expire events and (re-)compute aggregations, but on the other hand, aggregations are always accurate.

\subsection{Fraud-detection requires MAD systems}
\label{sec:realtimewindows}
Fraud-detection systems are responsible for, e.g., approve or block transactions, or raise alarms when money laundering is suspected.

As subcomponents of financial ecosystems, fraud-detection systems have very strict and demanding requirements, including all the MAD requirements defined above. It is easy to see why MD are required. Good customer experience implies a service that replies almost instantaneously (i.e., with sub 250ms latency at the 99.9\% percentile, \textbf{M}), and which is available at all times even when processing several thousand of requests per second~\cite{alibaba-cloud} (i.e., that is scalable and resilient to high-loads and failures, by being distributed, \textbf{D}).

Let's address accuracy (\textbf{A}) of sliding window aggregations. To make decisions, modern fraud-detection systems use machine learning models and rule based-systems, both fueled by streaming aggregations~\cite{kdd-Branco-deeplearning}.
For instance, queries such as \verb|Q1| and \verb|Q2| below can be used to \emph{profile} the common behaviors of card holders or merchants, and detect suspicious behavior.

Profiles computed over hopping windows are weaker as they are vulnerable to adversary attacks. Sophisticated fraudsters use many techniques to understand the best possible timings, and exploit attacks to occur at specific times, or follow a specific cadence, taking advantage of the predictable hop size. 
\begin{lstlisting}[
           language=SQL,
           showspaces=false,
           basicstyle=\ttfamily\small,
           commentstyle=\color{gray},
           label={lst:query},
           caption={Streaming 5-min metrics per card and merchant.},
           captionpos=b
        ]
Q1: SELECT SUM(amount), COUNT(*) FROM payments 
    GROUP BY cardId [RANGE 5 MINUTES]
Q2: SELECT AVG(amount) FROM payments
    GROUP BY merchantId [RANGE 5 MINUTES]
\end{lstlisting}

%

Additionally, profiles over hopping windows lead to inaccurate and counter intuitive results, compromising rule compliance -- either from internal bylaws, or from external regulators. 
As illustration, consider the following business rule: “\emph{if} the number of transactions of a card in the last 5 minutes is higher than 4, \emph{then} block the transaction”. If the window is implemented using 1-min hops, then the situation in Figure~\ref{fig:hop-example} can happen: the rule should trigger on the fifth event since it arrives within 5 minutes of the first one, but there is no hopping window including all 5 events in its boundaries using a 1-min hop.

%
To avoid this, one could argue that the hop could be adjusted to catch the intended behavior. However, the solution is not a panacea. First, the problem in Figure~\ref{fig:hop-example} can happen regardless of the hop size. Second, if the hop is reduced to a size where hopping windows behave almost like real-time sliding windows (e.g., 1-ms or even 1-sec step) then most stream processing engines systems crash or significantly degrade performance (cf. Section~\ref{subsection:flinkVsRailgun}).
This problem worsens with long windows. Fraud profiles use windows spanning over days, weeks, months and sometimes years. These include, e.g., the number of distinct addresses used in the last 6 months, or
the average user's expenditure of the past year.
As we shall see in Section~\ref{subsec:stateofart}, the performance of hopping windows depends on a ratio between the window size and the hop size. Hence, the longer windows are, the lower the precision must be, to achieve the same performance in terms of CPU, memory consumption and latency.


If streaming aggregations use only hopping windows and have metrics over long windows, then fraud systems need to use batch jobs and lambda-architectures~\cite{KiranMMDB15}.
In that case, imprecise but real-time aggregations are combined with precise but outdated aggregations over complex pipelines which are costly to maintain and hard to debug. Because of this, a lot of work has been committed to integrate batch and streaming in the same language and platform \cite{DBLP:journals/debu/CarboneKEMHT15, dataflow, trill}.
Nevertheless, in these systems, compliance is not achieved in real-time, limiting the possibility of preventing fraud from happening, and be restricted to use-cases where a post-mortem alarm is useful. Real-time 100\% compliance (i.e., accurate metrics per-event, \textbf{A}) is only possible using real-time sliding windows.

\begin{figure*}[ht!]
    \centering
      \includegraphics[width=0.90\textwidth]{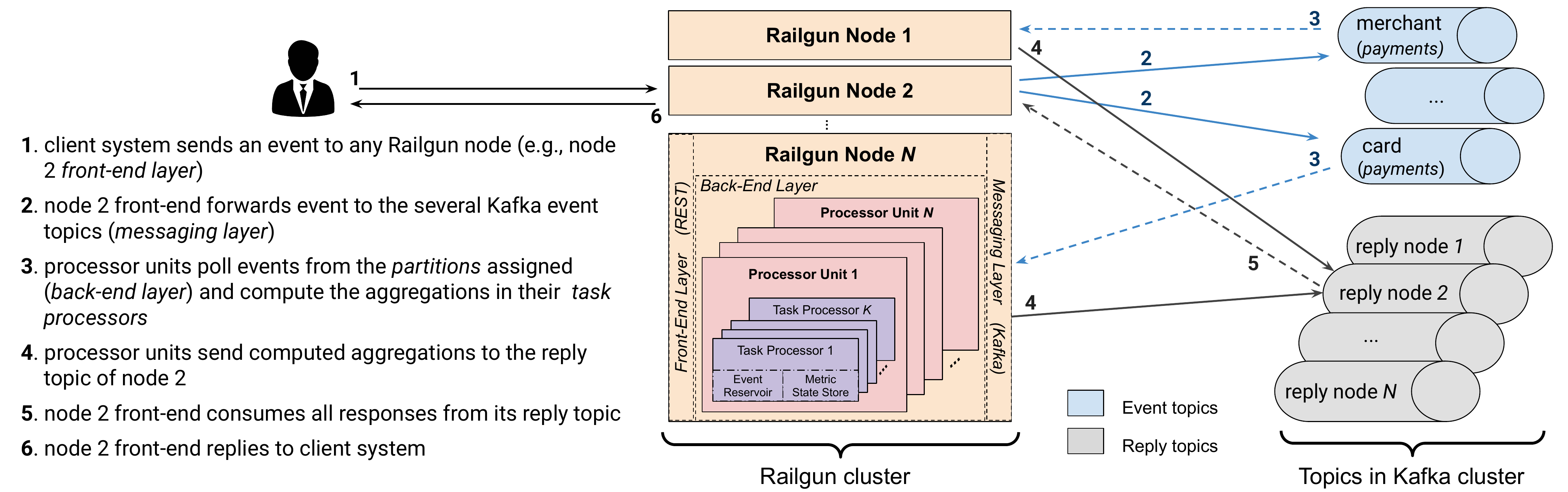}
    \caption{Tale of an event traversing Railgun.}
    \label{fig:tale-event}
\end{figure*}

\subsection{Related Work}
\label{subsec:stateofart}

The first generation of stream processing engines has risen from the database community.
These include seminal systems such as STREAM~\cite{DBLP:conf/sigmod/ArasuBBDIRW03}, TelegraphCQ~\cite{DBLP:conf/cidr/ChandrasekaranDFHHKMRRS03}, NiagaraCQ~\cite{niagaracq}, Aurora~\cite{DBLP:conf/sigmod/AbadiCCCCEGHMRSSTXYZ03}, and later, Borealis~\cite{borealis}, Coral8~\cite{coral8}, Event Insights~\cite{Navas10}, Siddhi~\cite{suhothayan2011siddhi}, Spade~\cite{spade}, Trill~\cite{trill}, Truviso~\cite{FranklinKCLRT09}, and SAP ESP~\cite{ZhangVDH17}. While a few of these systems implement real-time sliding windows~\cite{DBLP:conf/sigmod/ArasuBBDIRW03, niagaracq, trill, suhothayan2011siddhi}, almost all of them run as a single-node (Borealis and Spade are noteworthy exceptions), and they do not address the aggressive low-latency requirements we have. 

On the other hand, the latest generation of stream processing engines focus on processing high throughputs under low latency requirements, by building scalable, fault-tolerant, distributed systems. Examples of these include Flink~\cite{DBLP:journals/debu/CarboneKEMHT15}, Kafka Streaming~\cite{kreps2016introducing}, Spark Streaming~\cite{sparkstreaming}, and others~\cite{Akka, noghabi2017samza, ramasamy2019unifying, samzaVLDB17, stormSIGMOD14, heronSIGMOD15}.
However, to achieve high levels of performance, these systems need to limit what windows are possible. Neither Flink, Kafka Streaming, Spark Streaming, or to the best of our knowledge, any known, distributed streaming engine implements real-time sliding windows, restricting the window slide movement to fixed hops. 

So why are hopping windows so largely used?
Since the window size and hop size do not change during run-time, the number of physical window states active at any given time is fixed and exactly $\frac{\texttt{windowSize}}{\texttt{hopSize}}$. This allows streaming engines to do important optimizations, such as \emph{avoiding} to store events. Since the number of active window states is fixed, arriving events can be discarded once their contribution has been applied to all the active windows states. Hence, besides saving storage, these solutions also avoid processing event expiration.
As an example, recall \verb|Q1| and Figure~\ref{fig:hop-example}: the \verb|sum| and \verb|count| payments made by a card in the last 5 minutes with 1-min hop. Any event for this window affects 5 window states, and in this case, two variables~\footnote{one variable for the \texttt{sum} and another for the \texttt{count}} per window state. Every minute and, for every card active in the last 5 minutes, two new variables are created and the oldest two, expired.

Since their memory requirements are independent of throughput, this property makes hopping windows interesting as long as the ratio $\frac{\texttt{windowSize}}{\texttt{hopSize}}$ is low. When the ratio is higher, hopping windows bring extra problems with respect to latency, CPU usage and state scalability. 
For instance, a 60-min window with a 5-min hop implies 12 \emph{active} window states per metric; but if the hop is decreased to 1 second, for the same window size, the number of active window states becomes 3600. All of which must be updated per arriving event, and per metric. The situation becomes unsustainable especially on large windows ranging hours or days, unless the hop is adjusted accordingly, with severe consequences on the usefulness of the aggregations computed.
The impact of using small hops on large windows is further explored in Section~ \ref{subsection:flinkVsRailgun}.

Despite these drawbacks, hopping windows' wide usage and characteristics have driven substantial research and optimizations such as Cutty~\cite{DBLP:conf/cikm/CarboneTKHM16}, Scotty~\cite{DBLP:conf/icde/TraubGCBKRM18} and others~\cite{DBLP:conf/sigmod/TheodorakisKPP20, DBLP:conf/edbt/BensonGZMR20} 
that contribute to its popularity by delivering reduced hardware costs, support for out-of-order events, and distributed computations of a single query.

%

Unlike hopping windows, real-time sliding windows cannot discard events and therefore require storing and accessing events. For each event, metrics must be updated with the events exiting and entering the window. Since the window slides for every event, and not just at a specific hop, the optimizations discussed above for hopping windows are no longer possible.

Flink acknowledges the issue of high-precision metrics over time windows for low latency fraud-detection, with a customized solution~\cite{flink-fraud}: for each event, the solution computes each aggregation from scratch by iterating over all stored events (persisted in RocksDB) for those matching the window interval. 
This approach has quadratic performance, and since Flink was not designed to store events and manage event expiration, few optimizations are possible and performance degrades with long windows.
Hence, this solution has much worse performance than Flink's standard hopping windows (compared in Section~\ref{subsection:flinkVsRailgun}), failing our \textbf{M} requirement.


The interested reader can further explore related issues with stream processing engines elsewhere~\cite{BabcockBDMW02, GolabO03, HirzelSSGG13}.

\section{Railgun}
\label{sec:railgun}

Railgun is the paper's main contribution, and takes different design decisions when compared to the alternatives above: 1) works with real-time sliding windows to achieve aggregation correctness at all times (and not just at every hop); 2) uses an \emph{event reservoir} to efficiently store and access events under low latency and optimal memory usage; 3) manages an embedded aggregation store (persisted in RocksDB) for holding aggregation states and auxiliary data; and 4) takes advantage of a messaging layer (using Kafka) for distributed processing, fault-tolerance and recovery. 

The event reservoir, described in Section~\ref{subsec:reservoir}, exploits the predictable, time access pattern of events to optimize transfers between memory and stable storage, accessing nearly all events from memory using an eager caching. 
Plus, by optimizing the computation and storage of aggregation states, Railgun can deliver accurate results, per-event, with low latency.
This event reservoir is an evolution of previous work~\cite{slidem}. Here, we use a locally-attached storage in each Railgun node to minimize latency, a schema registry to support event schema evolution, and define a data format and compression for efficient storage, both in terms of deserialization time and size.

To distribute work and achieve scalability, Railgun uses Kafka topics, further split into \tp{}. Each stream can have multiple topics, depending on the combination of the metrics' \emph{group by}s, and each topic has multiple partitions which are distributed among the several nodes' processor units.
A \tp{} maps to a \emph{task} in Railgun, and is its minimal unit of work. Inside a task, we compute all aggregation metrics for a data stream subset, following a task \emph{plan} optimized to reuse computations.
To support high-availability and fault-tolerance, tasks have multiple \emph{replicas}, and a Railgun node processor concurrently handles a set of active tasks and a set of replica tasks. In addition, Railgun relies on Kafka's consumer group guarantees to ensure that tasks are always assigned to nodes, and provides a custom rebalance strategy (see Section~\ref{subsec:fault-tolerance}) to optimize task recovery, while safeguarding a balanced assignment from tasks to nodes. 

Railgun delivers event-by-event accurate results, by supporting real-time sliding windows, while still providing millisecond-level latencies at high percentiles. 
As we shall see in Section~\ref{sec:experiments}, Railgun can preserve these consistent tail latency results even when its scaled to achieve throughputs of one million events per second, and its performance is independent of the window size.

Railgun's high-level architecture is presented in Figure~\ref{fig:tale-event}, showing what happens when an event traverses the system. 
At this stage, and to simplify development, all Railgun nodes are equal and composed by layers: a \emph{front-end} layer to communicate with the client; a \emph{back-end} layer to compute aggregations and access storage; and a \emph{messaging} layer to handle distribution of tasks, detect failures, and communication between processor unit workers. This design could be revised in the future, with different nodes split by function. 

\subsection{Front-End Layer}
The front-end layer is the entry point for client requests, including events, requests for new metrics/streams, or deletions. 
Besides communicating with the client, the front-end layer distributes events and manages the overall cluster state (both using Kafka).

When a new stream is registered by the client, the front-end creates a set of partitioned topics to support it. The number of topics needed per stream depends on the number of distinct \emph{group by} fields of the stream. 
As further detailed in Section~\ref{sec:computation}, a stream is mapped to one or more topics to support work distribution across the several processing units. Hence, when a new event arrives (step 1 of Figure~\ref{fig:tale-event}), it is the front-end layer responsibility to route events to all of its topics (step 2). E.g., in case of Example~\ref{lst:query} and Figure~\ref{fig:tale-event}, to simultaneously publish any event of stream \verb|payments| to topics \verb|merchant| and \verb|card|, as they are \emph{group by} aggregations for both \verb|merchantId| and \verb|cardId| (\verb|Q1| and \verb|Q2| of Example~\ref{lst:query}).

The metrics of a stream are computed by one or more back-end instances possibly residing in different Railgun nodes (step 3), which reply to the node originally posting the event in its dedicated reply topic (step 4). The front-end is also in charge of collecting the several computations (step 5) from its reply topic, and responding to the client with all the aggregations computed for that particular event in a single message (step 6).

\subsection{Back-End Layer}
\label{subsec:backend}
The back-end layer is responsible for the computation of metrics. 
Each back-end instance has one or more \emph{processor units}, each with its own dedicated thread. A processor unit manages a set of \emph{tasks}, all computed within a single thread, to reduce context switching and synchronization, thereby optimizing for latency.
%

Importantly, each processor unit is completely independent of each other, and two processor units deployed on the same physical Railgun node are logically equivalent to two Railgun nodes with one processor unit each. As such, by having many processor units inside a single node, we can exploit multi-core machines efficiently.
Notwithstanding, and as we shall see in Section~\ref{subsec:fault-tolerance}, distributing processor units among multiple physical nodes can bring advantages in terms of fault-tolerance, work rebalance and high availability. 

%
\begin{algorithm}
	\caption{Processor Unit Logical Loop} 
	\label{algo:processor}
	\begin{algorithmic}[1]
		\While {$running$}
		    \State $processOperationalRequests(requests)$
		    
			\State $activeMessages \leftarrow consumerActiveTasks.poll()$
			
			\State $replicaMessages \leftarrow consumerReplicaTasks.poll()$
			
			\For {$message \leftarrow activeMessages \cup replicaMessages$}
			    \State $t \leftarrow message.topic$
			    \State $p \leftarrow message.partition$
			    \State $taskProcessor \leftarrow taskProcessors.get(t, p)$
                \State $answer \leftarrow taskProcessor.processMessage(message)$
                \If{ $(message \in activeMessages)$}
                    \State $sendReply(answer)$
                \EndIf
			\EndFor
		\EndWhile
	\end{algorithmic} 
\end{algorithm}
The processor unit duties are listed in Algorithm~\ref{algo:processor}. While running, the processor handles operational requests (such as adding/removing new streams or metrics); consumes message events for its assigned (active and replica) \emph{tasks}; forwards the events to their appropriate \emph{task processors} which handle event storage and task computation; and replies with the computation answer to a dedicated reply topic, for active tasks. Each processor unit is then both a consumer of event topics (inbound stream events), and a producer for reply topics (outbound aggregation results).

A task encompasses the computation of all metrics associated with a given \tp. A \tp{} is the unit of work distribution among nodes and processor units. As we shall see in Section~\ref{subsec:messaging}, events are routed, consumed and processed according to their \tp.
Processor units have \emph{active} tasks, for which they are the leaders, and \emph{replica} tasks for which they are hot standbys. To poll messages from the messaging layer, the processor unit has two consumers, for each type of task. Separating the consumers allows us to prioritize active tasks, and better exploit Kafka rebalance protocol and consumer group guarantees (cf. Section~\ref{subsec:fault-tolerance}).

Finally, processing message events and computing metrics for any task happens within a \emph{task processor}. Each processor unit has as many task processors as (active or replica) tasks it has assigned, all computed within a single thread. Thus, while the number of processors units sets the cluster's level of parallelism, the number of task processors in Railgun establishes the cluster's level of concurrency.

\subsection{Messaging Layer}
\label{subsec:messaging}
The goal of the messaging layer is many-fold: 1) serve as the communication layer between different Railgun nodes, including between the front-end and back-end layer of the same physical node; 2) detect Railgun node failures, during the polling of messages; 3) support the recovery of Railgun node failures, by reliably storing events and aggregation replies which can be rewinded upon request. 

Currently, Railgun uses Kafka~\cite{kreps2011kafka}. Kafka is a distributed, highly scalable and fault tolerant messaging system, with high throughput and low latency guarantees. In opposition to push-based systems such as RabbitMQ~\cite{rabbitmq}, Kafka follows a pull-based approach where consumers continuously poll for new messages by providing their individual offset since the last poll. This is important since it allows a Railgun node to recover by rewinding the stream and replaying unprocessed messages without degrading the end-to-end latency of the overall system.
Kafka stores messages in \emph{topics} and provides built-in capabilities to split topics into several \emph{partitions}, for higher throughput and parallelism. As described further in Section~\ref{sec:computation}, Railgun takes advantage of Kafka's partitions to distribute work among the several Railgun processor units and their task processors.

Except with the client, all communication happens using Kafka, and Railgun nodes are, simultaneously, Kafka producers and consumers. Railgun nodes communicate for many reasons: 1) to broadcast operational requests triggered by the client, such as creating/deleting a stream, metric or partitioner; 2) to propagate events to all of their topics; 3) to share aggregation partial results which are collected before answering the client; 3) to handle cluster maintenance tasks such as node removal/addition and the corresponding rebalance of tasks in the cluster.

Although we could have chosen any other messaging system (as long as it implements the same pull-based and partition concepts),
we choose Kafka due to its proven performance, and for providing us with many features that simplify our development.
First, Kafka is actively monitoring what consumers enter or leave the cluster, requiring every consumer to send heartbeats periodically and assuming consumer failure, otherwise. Therefore, whenever the consumer landscape changes, Kafka detects this and, at step 3 of Algorithm~\ref{algo:processor}, triggers a callback to rebalance the cluster. At this moment, Railgun's assignment strategy (described in Section~\ref{subsec:fault-tolerance}) takes over, e.g., to decide how to reassign the multiple \tp{} -- or tasks -- of a failed node. 

Second, in Kafka, consumers can be organized in \emph{consumer groups}, to allow load distribution among a group of several consumers, with important guarantees. Namely, by design, Kafka ensures that there is exactly one consumer within a consumer group subscribing and consuming messages from a \tp. We take advantage of this property to ensure that there is exactly one processor actively responsible for a task, by configuring all Railgun active task consumers to belong to the same consumer group.
On the other hand, replica task consumers all have different consumer groups to enable multiple Railgun processors to subscribe to the same \tp. Therefore, we ensure high-availability by having multiple hot replicas available in the cluster. 


%
Finally, we ensure exactly-once semantics by combining Kafka's at-least-once guarantees, with an event deduplication logic on the application back-end layer  (see Section~\ref{subsec:reservoir}). 

%

\subsection{Railgun Operators}
\label{subsec:grammar}
\begin{figure}[b]
    \centering
    \includegraphics[width=0.476\textwidth]{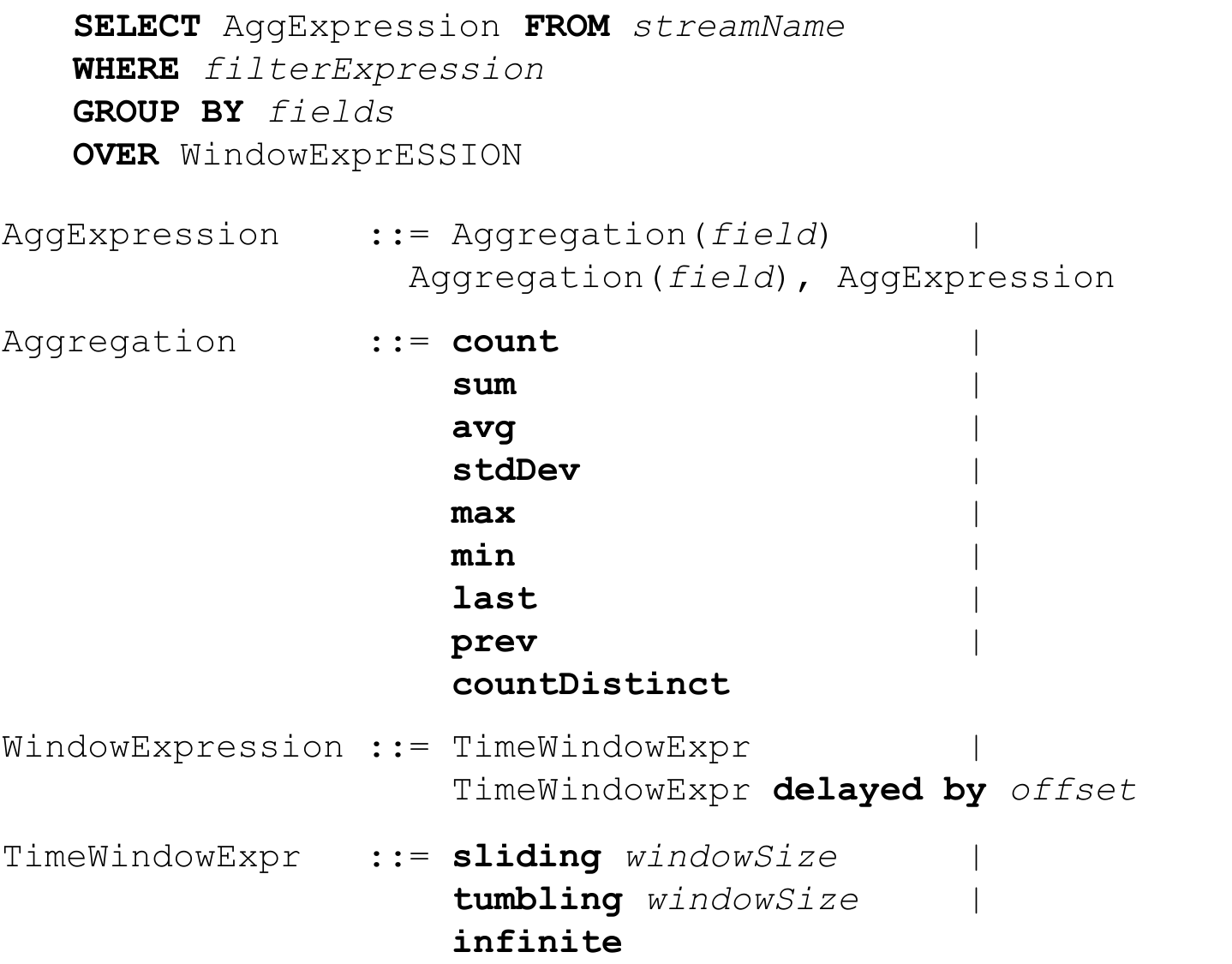}
    \caption{Railgun Operators.}
    \label{fig:grammar}
\end{figure}

Railgun's language is summarized on Figure~\ref{fig:grammar}. We support SQL-like query statements, where each statement can include multiple aggregations over a single stream.

Currently, Railgun does not natively support stream joins. In practice, we implement joins (e.g., between a stream and a lookup table) prior to the streaming engine, in an enrichment stage.

Besides sliding and tumbling windows, we support infinite windows, i.e., windows where events never expire (e.g., the count of all distinct addresses of a client). Any window can be delayed, i.e., where instead of considering the window against the latest arriving event, we can delay its starting by a specific delay offset. Delayed windows are especially useful in bot-attacks scenarios.
%
We choose not to support hopping windows, since we see them as an approximation of our sliding windows. In practice, we never found a use-case where hopping windows are functionally preferable to sliding windows.
Although, at this time, we only support time windows, the system could be easily extended to support windows whose size is based on the number of event. 

Finally, we use jexel expressions~\cite{jexel} as our filter expression language to support additional flexibility, using Java.

\section{Computation and Distribution}
\label{sec:computation}
The \emph{(topic, partition)} combinations affect how work is distributed among the several Railgun nodes, and processor units.
Each data stream has a topic for each configured top-level entity, which we call \emph{partitioner}.
For instance, Figure~\ref{fig:tale-event} shows two topics for two partitioners - \texttt{merchant} and \texttt{card} - over the same \texttt{payments} stream (corresponding to the \emph{group by}s for \verb|merchantId| and \verb|cardId| of Example~\ref{lst:query}).
Currently, the set of partitioners is manually provided by an administrator when a stream is created, depending on the possible \emph{group by}s of the metrics. Since computation is contained within a task processor, to provide accurate metrics, we need to ensure that whenever a task processor is computing a metric for an entity (e.g., of a particular card, or merchant), it receives all that entity's events.
As a result, metrics with multiple \emph{group by}s over a stream, as in \verb|Q1| and \verb|Q2| of Example~\ref{lst:query}, may cause the event to be forwarded to more than one topic (step 2 of Figure~\ref{fig:tale-event}). Events are, in fact, replicated as many times as the number of partitioners (i.e., top-level \emph{group by}s such as \verb|cardId| or \verb|merchantId|) needed for a stream, resulting in a few topics per stream.

Notwithstanding, the number of topics needed is usually small, and it is not necessarily equal to the number of distinct group by keys of all stream metrics defined (which could lead to dozens of topics). Accurate metrics, only need events to be hashed by a \emph{subset} of their group by keys.
E.g., two metrics, one grouping by \texttt{card} and \texttt{merchant}, and the other by \texttt{card}, could both use topic \texttt{card}. This reduces Kafka's storage, and thus the front-end receives, from configuration, the partitioners for a given a stream upon stream creation. 
Partitioners can also be set after a stream is created, but this causes the creation of new \tp{} and a consequent cluster rebalance, which can be an expensive operation. However, and as we shall see in Section~\ref{subsec:fault-tolerance}, our rebalance strategy is sticky, i.e., it preserves task assignment to their previous processor as much as possible. As a result, the processing of the existing \tp{} of the cluster is generally unaffected when a rebalance is triggered for adding new topics. Plus, adding a new partitioner is done only when a new top-level \emph{group by} is needed which, in practice, is rarely required after a stream is created.

Finally, a \emph{partition} is a Kafka concept that further allows us to distribute work among several consumers (i.e., processor units). 
For partitioning, Kafka allows producers to provide a key when publishing a message, which is hashed according to the number of partitions defined for a topic.
When a key is provided, it is guaranteed that messages with the same key will always be delivered to the same \tp. In Railgun, we configure the message key for each topic to be the \emph{partitioner}. When a new event arrives for a stream, the front-end layer node receiving the event publishes as many messages as partitioners defined for that stream.

The number of partitions for each topic is defined according to the expected load of each stream-partitioner. 
Recall that the \tp{} is the minimal work unit, and the distinct number of \tp{} establishes the number of task processors created in Railgun, where each task processor handles a single pair of \tp{}. Hence, by increasing the number of partitions, we increase the cluster's level of concurrency.
As described, by exploiting Kafka's guarantees over consumer groups, we ensure there is exactly one \emph{active} task processor for each existing \emph{(topic, partition)}. However, to support high-availability, the number of task processors is multiplied by the replication factor. If there are $n$ distinct \emph{(topic, partition)}, and $r$ is the replication factor, there are exactly $n \times r$ task processors working in the cluster.

\subsection{Task Processors}
The computation of \emph{all} metrics for a given \emph{(topic, partition)} encapsulates a \emph{task}, which is done within a \emph{task processor}. Each task processor is designed to share nothing, and work independently of other task processors, without the need to synchronize or access shared storage. To support this, each task processor is further composed of: an \textit{event reservoir} that stores its own events; a \textit{state store} holding aggregation states of each configured metric; and an execution task \textit{plan} -- i.e., a directed acyclic graph (DAG) defining how metrics will be executed.

\subsubsection{Event Reservoir}
\label{subsec:reservoir}
The event reservoir is a structure that stores all the events of a task processor, and allows efficient access of the events as they are needed by windows to update the aggregations. The event reservoir is an evolution of previous work~\cite{slidem}, and has two parts: a very small memory part holding the tail and head of each window, and a potentially large part stored in the node's local disk holding the full set of events.

Processing an event starts with the event reservoir, where events are persisted to and loaded from disk as needed.
Before persistence, events are serialized and compressed into groups of contiguous \emph{chunks}. Grouping events into chunks helps to reduce the number of I/O operations needed. In a reservoir, all I/O operations are asynchronous, to not affect event processing latency. Chunks hold multiple events and are kept in-memory until they reach a fixed size, after which they are \emph{closed}, serialized, compressed, and persisted to disk over \emph{ordered} and append-only files. Similarly, files hold multiple chunks of events, until they reach a fixed sized, after which they become \emph{immutable}. Since files are immutable and events follow a monotonic order given by their timestamp, we can efficiently support random reads by maintaining an auxiliary index in-memory, from timestamps to files. Supporting random reads is especially useful when adding metrics with new windows to the system. 

Since chunks are frequently persisted to disk, recovery is simplified, as only the most recent events can be lost, and quickly recovered from Kafka broker nodes.

Out-of-order events are supported until the closure of a chunk, i.e., as long as the event timestamp occurs after the last closed chunk timestamp. After that moment, and depending on the configuration, events are either discarded, or have their timestamp rewritten to the first timestamp of the chunk.
For scenarios requiring extensive support for out-of-orders events, we can delay the chunk closure by a time period provided by configuration. This keeps chunks in a transition state in-memory for a threshold period, on which they are closed for recent event, but are still open for late events. In a way, this configuration can be seen as a watermark~\cite{watermarks}, which should be used with care, as it may cause an increased memory consumption, and recovery delays. However, since latency is a prime goal of our system, we might delay the closure of a chunk, but we never delay the answer and computation of event metrics, as opposed to systems such as Spark Streaming or Flink. 
Events are also deduplicated based on an \emph{id}, against the chunks still in-memory, to avoid processing an event more than once.

The reservoir takes advantage of the predictable event consumption pattern in stream processing where events are \emph{always} consumed by their timestamp order, by advancing windows. Namely, the reservoir provides very efficient \emph{iterators} which transparently load chunks of events into memory as they are needed by windows. Iterators eagerly load adjacent chunks into cache when a new chunk is loaded from disk, and starts to be iterated. Hence, when a window needs events from the next chunk, the chunk is normally already available for iteration. Notwithstanding, if for some reason, chunks are evicted from the application cache right before they are requested, thus resulting in \emph{syscall} to fetch them from disk, the request will likely not trigger an actual read request to disk. 
Since chunks are organized as a sequence in a file, the operating system I/O will likely already read ahead the chunk contents into page cache. Thus, when a chunk not in cache is requested, it is likely delivered from the OS page cache, paying only the deserialization cost -- a fraction of what it would be if an actual I/O request to disk was required.
This predictability helps us relax the hardware demands for the reservoir tremendously, as even for low latency scenarios, we can use a network-attached storage or HDDs, instead of holding all events in memory, which significantly reduces the total cost of ownership.
\begin{figure}[t]
    \centering
    \includegraphics[width=0.48\textwidth]{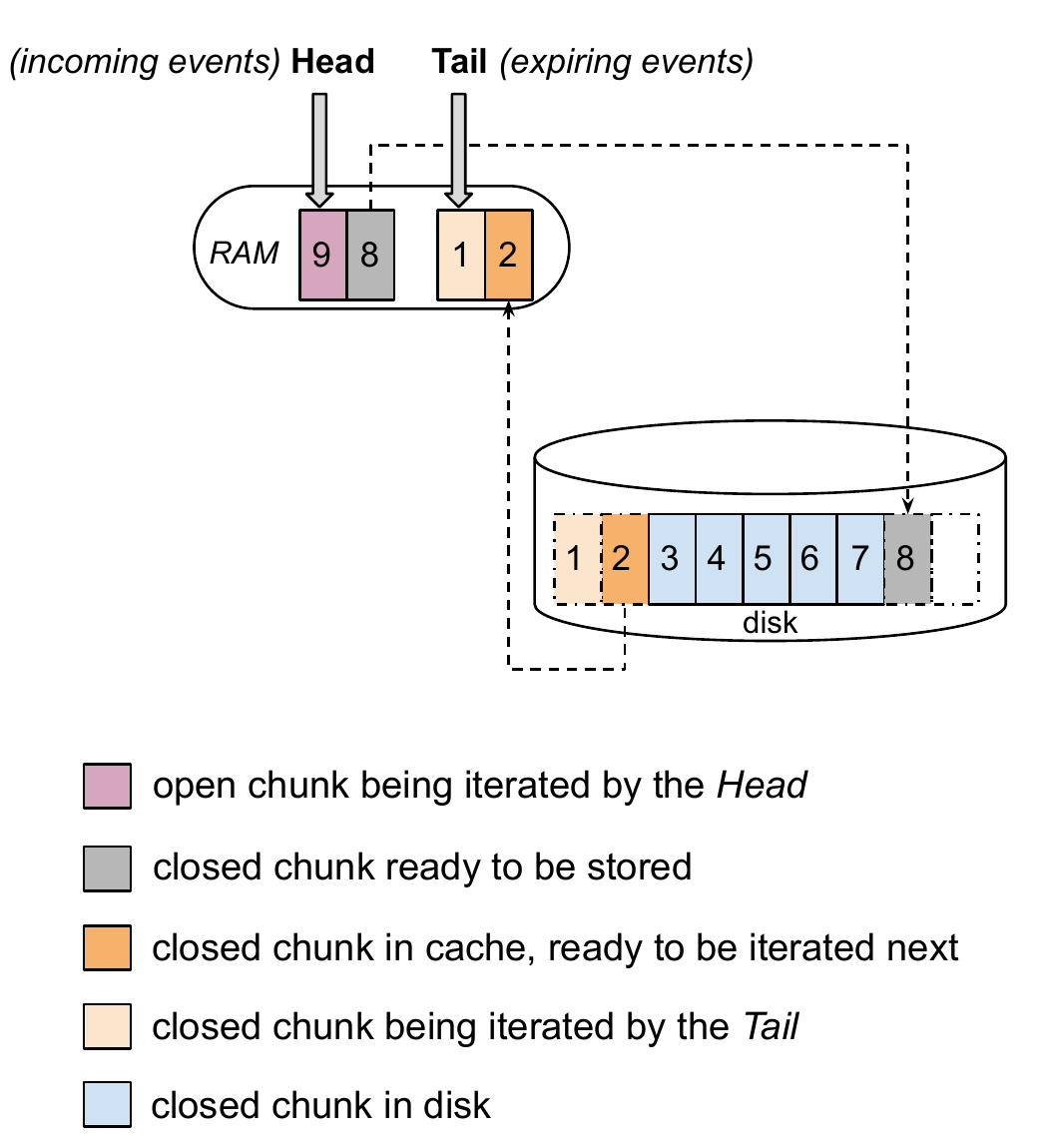}
    \caption{Iterators for a Window in Event Reservoir.}
    \label{fig:reservoir}
\end{figure}

Along with a reservoir, we keep a \emph{Schema Registry} to support schema evolution of events. Before persisted, chunks are serialized using a specific events' schema and stored referencing their current schema id. Each time the event schema changes, a new entry is added to the schema registry, and the current schema id reference is updated. Whenever we need to deserialize a chunk with an old schema, we just retrieve it from the schema registry. 
Chunks are also compressed aggressively to guarantee a good compression ratio. This is important to minimize storage overhead, since events can be replicated across multiple task processors. 

Regardless of the window type and window size, only a tiny fraction of events need to be kept in-memory, as illustrated by Figure~\ref{fig:reservoir}. By default, each window has two iterators -- one for the head of the window (incoming events), and another for the tail (expiring events) -- and each iterator only needs one chunk in-memory\footnote{However, due to eager caching, more chunks may be in-memory.}. Whenever possible, we reuse iterators among windows. For instance, over the same reservoir, two real-time sliding windows always share the same head iterator (e.g., a 1-min and a 5-min sliding window share the same head iterator, which points to the most recently arrived event). 
%
This design makes the reservoir optimal for I/O~\cite{DBLP:journals/ipl/Roy07}, and extremely efficient for long windows. Namely, and except for the extra storage needed (minimized by compression and serialization), windows of years are equivalent to windows of seconds -- in performance, accuracy, and memory consumption.

\subsubsection{Task Plan}
The task plan is a DAG of operations that compute all the metrics of a task, following the order: \texttt{Window -> Filter -> Group By -> Aggregator}. Since
we often have metrics sharing the same \texttt{Window}, \texttt{Filter}, and \texttt{Group By} operators, the plan optimizes these by reusing the DAG's prefix path.

\begin{figure}[hb]
    \centering
    \includegraphics[width=0.45\textwidth]{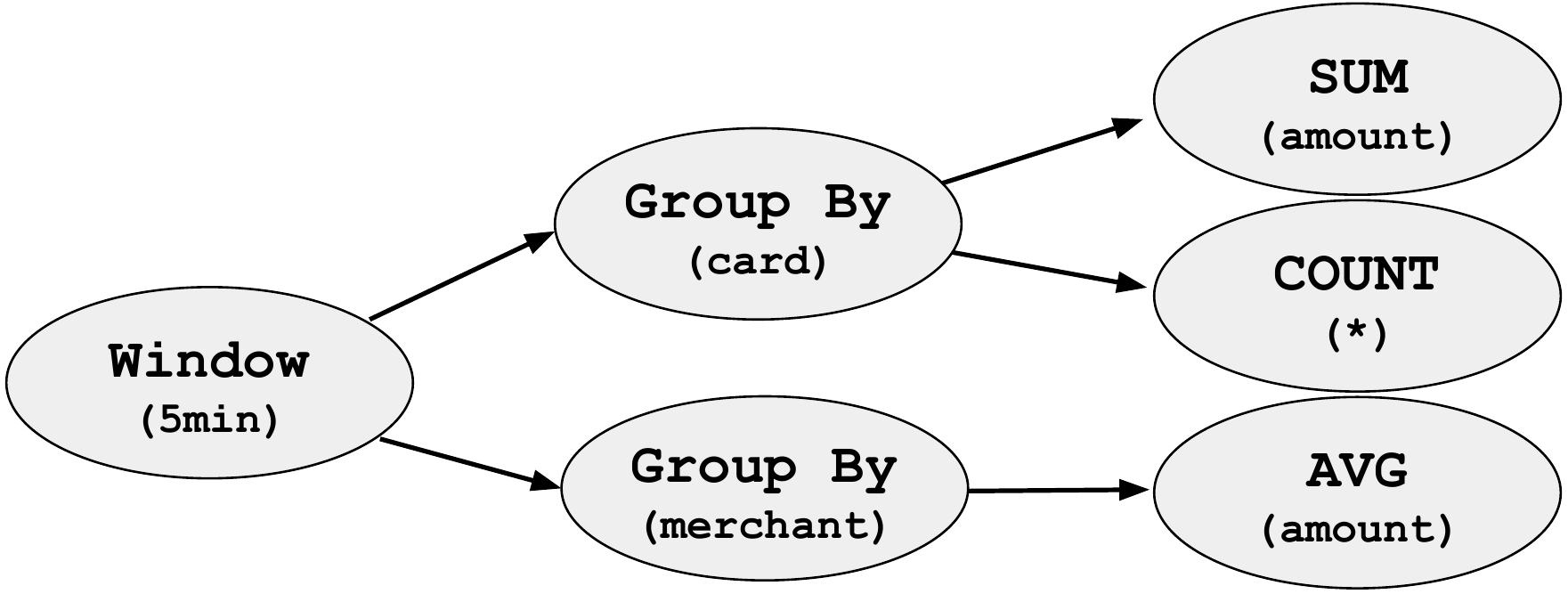}
    \caption{Plan DAG of Example~\ref{lst:query}.}
    \label{fig:dag}
\end{figure}

Figure~\ref{fig:dag} shows the DAG of Example~\ref{lst:query}. In it, all metrics share the same window, but \verb|Q1| groups by field \texttt{card} while \verb|Q2| by field \texttt{merchant}.
Optimizing the DAG to reuse operators prevents us from repeating unnecessary computations, especially ones related with windows. Every time a plan advances time, the \texttt{Window} operator produces the events that \emph{arrive} and \emph{expire}, to the downstream operators of the DAG. 
However, to make these optimizations, we restrict Railgun's query expressibility to follow a strict order of operations, defined in Section~\ref{subsec:grammar}. This is in contrast to general solutions such as Flink or Spark Streaming, which provide a more flexible API, harder to optimize.

While the roots of the DAG iterate over the reservoir and push events downstream, the leafs (i.e., \texttt{Aggregator} operators) use the \textit{state store} to keep and access the results of the aggregations. 

\subsubsection{State Store}
Similar to other streaming engines such as Flink, Railgun uses RocksDB~\cite{rocksdb} to store, for each metric key value, the latest aggregations results and auxiliary data. Built on top of LSM-trees~\cite{ONeilCGO96}, RocksDB has proven to be a reliable, memory efficient and low latency embedded key-value store.

The amount of RocksDB keys, and their access pattern is tightly related with the task plan. Namely, each key represents a particular metric entity in a plan, and the amount of keys accessed per event match the number of DAG's leaves of a plan. For instance, for each event, the plan of  Figure~\ref{fig:dag} will access exactly two keys for the card aggregations (\verb|sum| and \verb|count|), and one key for the merchant (\verb|amount|). 
Each key holds the aggregation current value for the specific window and the specific entity. Depending on the aggregation type, auxiliary data might be stored with the aggregation. For instance, an \verb|average| requires storing also a counter, while a \verb|sum| or a \verb|count|, do not require any extra data other than the current value. Additionally, \verb|max| and \verb|min| store a deque structure~\cite{donaldknuth}, the \verb|stdDev| stores the three parameter to compute the Welford's online algorithm~\cite{Welford62noteon}, whereas the \verb|countDistinct| uses an auxiliary column-family in RocksDB to hold the counts.

To support fault-tolerance, RocksDB provides checkpointing, which forces the flushing of all data in-memory to disk. However, by design, even without checkpoints, RocksDB data is only kept in-memory for a short period of time, and is frequently persisted to disk. This makes checkpoints very efficient, since only a small amount of data needs to be written to disk, at a given time.
We synchronize checkpoint triggers among the event reservoir and the state store, and references to the latest event checkpoint offset of each task processor and node are frequently stored in a dedicated Kafka topic, which allows us to ensure that both stores can be easily recovered during a failure. 

\subsection{Rebalance and Recovery}
\label{subsec:fault-tolerance}

The assignment of tasks to nodes and processor units is triggered
during a Kafka rebalance, which happens whenever nodes or tasks are added/removed from the cluster.

As previously mentioned, Kafka tracks consumers within each consumer group to guarantee load distribution and message delivery. Kafka consumer group protocol ensures each \tp{} has exactly one consumer assigned in a group. In particular, 
it is impossible to have a \tp{} assigned to multiple consumers of the same group, and if there are more consumers in a group than \tp{} combinations, a consumer might not have any \tp{} assigned.
To achieve this, Kafka is continually tracking what consumers are registered for a consumer group, and is actively receiving heartbeats for each consumer.
When a consumer enters or leaves (either due to a failure or graceful shutdown) a Kafka consumer group, a rebalance happens.

\begin{figure}[b]
    \centering
    \includegraphics[width=0.475\textwidth]{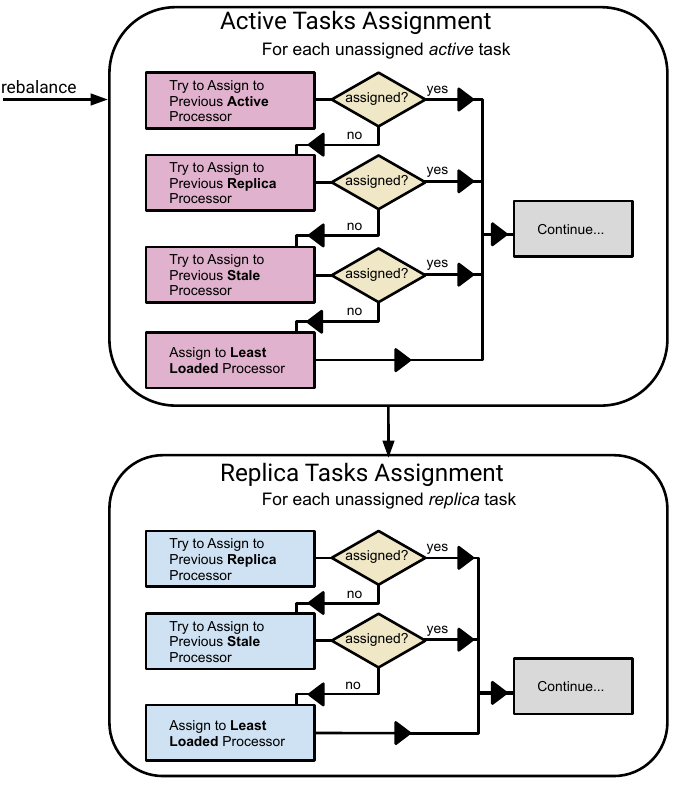}
    \caption{Railgun Sticky Assignment Strategy.}
    \label{fig:stickyAssignment}
\end{figure}

When a rebalance is triggered, one of the Railgun nodes (the consumer group coordinator) decides how \tp{} pairs are distributed among each consumer. While Kafka makes available several different strategies to assign \tp{} to consumers, Railgun uses a custom assignment strategy, built upon Kafka's sticky assignment implementation.
The assignment strategy logic is shown in Figure~\ref{fig:stickyAssignment} and is split into two main assignments: \emph{active} tasks, and \emph{replica} tasks.

Recall that a task maps to a specific \tp, and that consumers (located within processor units) can have tasks assigned as active and as replicas. This distinction between active and replica tasks has more to do with Kafka consumer groups, than with computation. A processor unit will process messages from both active and replicas, and compute their aggregations in the same way. The only difference is that messages from replica tasks will not trigger a response from the processor, since this responsibility is exclusive of active consumers.
While active consumers share the same consumer group, replica consumers have different consumer groups from active and other replica consumers. This allows us to assign a \tp{} to a single active consumer in the cluster, but also to multiple replica consumers, simultaneously. Since both active and replicas tasks consume messages from the same \tp, they always consume them on the same order, ensuring consistency on the reservoir and metric state store for the several replicas.

Achieving a perfect assignment of tasks is an NP-complete problem as it implies solving a search over multiple-goals~\cite{DBLP:conf/coco/Karp72}: 1) minimize recovery time by minimizing data shuffling; 2) respect the cluster balance for active/replica tasks; 3) distribute the load fairly among the processors.
In practice, our assignment strategy logic, shown in Figure~\ref{fig:stickyAssignment}, implements a greedy approach that always protects two invariants: 1) tasks are only assigned to a physical node once; 2) the load respects a predefined processor \emph{budget}.

The first invariant aims to avoid the loss of multiple task copies when a node fails, or it is decomissioned.
Hence, while metric computation within each task processor is agnostic to where processor units are located, the assignment strategy is not. Consequently, the strategy takes as input the locality of each processor, to ensure that a physical node will never be assigned the same task twice during the same a rebalance iteration assignment.

%
Another constraint relates with how load is distributed among the several consumers. To ensure that load is fairly distributed among the cluster, for each assignment, the strategy sets the maximum \emph{budget} of each processor unit as: $budget = \frac{tasks}{processor \; units}$. Each time a rebalance is triggered, the available budget of a node is reset to this value. Whenever a task is assigned to a processor, the available budget of a node is decremented by 1. When its budget reaches 0, the processor can no longer receive assignments. At this stage, we consider all tasks as equal, however, in the future we might want to give tasks a different weight, depending on their computational cost (viz., partition load, event reservoir size, etc.).

To ensure these two invariants, upon each new rebalance iteration assignment, the group coordinator collects cluster metadata to understand how many tasks, physical nodes and processor units exist, and how processors are located within each physical node.

Regardless of the task's type, the goal of the sticky assignment strategy is to avoid data reshuffling as much as possible, while respecting the two invariants above. %
Therefore, the first step of the algorithm is to always try to maintain the task in the consumer that had the task in the previous iteration. An assignment might fail, if we violate one of the two invariants.

Active tasks that can not be kept in the same processor will be attemptedly assigned to a processor previously holding one of its previous replica tasks. If more than one processor replica is available for assignment, we choose the one with the least load.
If a task cannot be assigned to a processor replica (because of the conditions mentioned above), we will try to assign the task to a processor that still has the task as \emph{stale}. A stale task is a task for which the processor used to be assigned in the past (either as active or replica), but lost its assignment during a rebalance.
In other words, processors with stale tasks are processors that still have data ``leftovers'' available for that task. Hence, assigning a task to one of its stale processors, only requires recovering a \emph{subset} of the data, instead of the whole data. Again, in case of ties, we choose the processor with the least load. In the future, we might consider to give priority to the processor that has more data available (i.e., the processor that needs less data shuffle to recover).

Lastly, if none of the assignments is possible, we simply assign the task to the consumer with the most available budget. 

This assignment strategy, in combination with a replication factor, allows us to achieve high-availability. When a node fails and a rebalance is triggered, active tasks are always the first tasks assigned, maximizing the probability to be allocated in nodes already holding that task. In this case, the processor does not need to recover any data and the task is recovered immediate.

When a task is assigned to a processor that was not actively processing it before (either as active or replica), a recovery process happens within that processor, which might affect these tasks' immediate availability. However, since we prioritize the assignment of active tasks over replicas, this is extremely unlikely to happen for active tasks.
As usual, the replication factor is set according to the number of failures we want to tolerate before affecting a task's availability. In practice, in most of our deployments we use a replication factor of three. 

To perform recovery, the processor triggers a request to another processor unit that still has data available -- to copy the event reservoir, the state store, and the last event offset since its last checkpoint. After data is transferred, the processor starts its execution by consuming messages from Kafka since the last checkpointed offset.
Importantly, a processor with stale data, only needs to copy the delta between its own last checkpoint and the newest checkpoint available in the cluster, thereby minimizing the time to recover.

\section{Experiments}
\label{sec:experiments}
To validate our approach we present three experiments, in order to: 1) measure how Railgun's real-time sliding windows compare with the performance of Flink's hopping windows; 2) assess how Railgun's latencies are affected with different window sizes and a growing number of windows; 3) measure how Railgun can be scaled to handle more load by adding more nodes to the cluster. 

All of our experiments focus on tail latency, i.e., the latency of the system at the high percentiles of latency distribution. Tail latency measures how stable and consistent a system is, and it is one of our core Service Level Objectives (SLOs) contractualized with our clients. Addressing tail latency is hard for multiple reasons. First, tail latency is very hard to debug, as it can be affected by multiple types of resources, either individually or by its combination, including disk storage differences, garbage collector configurations, or network bandwith. Second, in a distributed setting, tail latency is impacted by its slowest component~\cite{DBLP:journals/cacm/DeanB13}, which in our case could mean the slowest Kafka broker, or the most loaded Railgun processor responsible for a larger data partition.

To make our experiments representative, in all three of them we used a real fraud dataset from one of our client. This dataset includes 103 fields, and with it, we aim to simulate real-world dictionary cardinalities for the aggregation states, and the expected load differences among the several Railgun processors. 

In all of our experiments we have one or more injectors producing events to a single Kafka input topic (for computing metrics over a single \emph{group by} over one stream) at a sustained throughput. Each computing node consumes events from the input topic, computes the several aggregations, and sends the results to the injector's dedicated reply topic. Latencies are measured by the injector based on the reply message time. I.e., we compute the \emph{end-to-end} latency since the injector sends the message to Kafka, until the moment it consumes from Kafka the aggregations response. As such, this latency includes the network time, the communication overhead using Kafka, and the processing time of the Railgun (or Flink) computing node. These latencies are corrected to take into account the coordination omission problem~\cite{CoordOmi}.

The first two experiments (Section~\ref{subsection:flinkVsRailgun} and Section~\ref{subsec:scalingRailgun}) aim to validate several of our design decisions, and have a simpler setup. For these, we use 3 \textit{m5.2xlarge} AWS instances, with 8 vCPUs, 32GB of RAM, using only EBS storage. We use Kubernetes to deploy 1 Kafka \emph{pod} (with Zookeeper), 1 injector, and 1 computing engine -- either Railgun or Flink (v1.11.0) -- with a JVM heap of 10GB all in separate VMs (by using Kubernetes anti-affinity rules). We use two Kafka topics -- one to publish events with 10 partitions; and another to consume responses with 1 partition. Since we only use one Kafka node, replication is set to 1. In both these experiments the throughput is fixed at 500 ev/sec. 

For the third experiment (Section~\ref{subsec:distributedRailgun}), we simulate a more realistic scenario with multiple Railgun nodes and Kafka brokers. For this, we use multiple \textit{m5.4xlarge} AWS instances, with 16 vCPUs, 64GB of RAM, with only EBS storage. Again, we use Kubernetes with anti-affinity rules to deploy multiple injector nodes, 30 Kafka brokers, and between 1 and 50 Railgun nodes - each with a JVM heap of 32GB - all in separate VMs. In all runs of this experiment, each Railgun node is configured with 8 processor units. Injectors publish events to a single topic, configured with a number of partitions that match the number of Railgun consumers of each run, viz., \# processor units $\times$ \# Railgun nodes. In this case, Kafka replication is set to 3, and the injectors are configured with \emph{ack=all} to ensure delivery guarantees.  We have one reply topic dedicated to each producer with 6 partitions. For aggregation replies, since they are usually discarded by the upstream systems if the reply arrives after our latency requirements, we set the topic replication to 1. In opposition to the previous two experiments, here we vary the injection throughput rate from a minimum of 25 thousand ev/sec to a maximum of 1 million ev/sec, according to the number of Railgun nodes in the cluster.

\subsection{Comparing Flink with Railgun}
\label{subsection:flinkVsRailgun}

On the first experiment we show the limitations stemming from using hopping windows, as described in Section~\ref{subsec:stateofart}, and how they compare with Railgun's real-time sliding window.
For this, we chose Flink as it is one of the most performant~\cite{KarimovRKSHM18, ChatterjeeM18} and widely-used stream processing systems, and the closest to our functional needs.

As mentioned above, we use a single-node deployment for this experiment. The goal is to more fairly compare Railgun with Flink, using a simple setup, where, under a sustained throughput of 500 ev/s, we compute a single metric -- the \verb|sum(amount)| per card over a 60-min window. 
In Railgun we use a 60-min real-time sliding window, while for Flink we use hopping windows, and vary their hop size from 5 minutes to 1 second. Our goal is to show how Flink latency distributions behave when we attempt to approximate hopping windows to sliding windows. 
All experiment runs are of 35 minutes, where the first 5 minutes are for warmup, and ignored for latency purposes.
To optimize Flink for latency rather than throughput, we set Flink's Kafka Client to use a batch timeout of 0.

\begin{figure}[h]
  \includegraphics[width=0.45\textwidth]{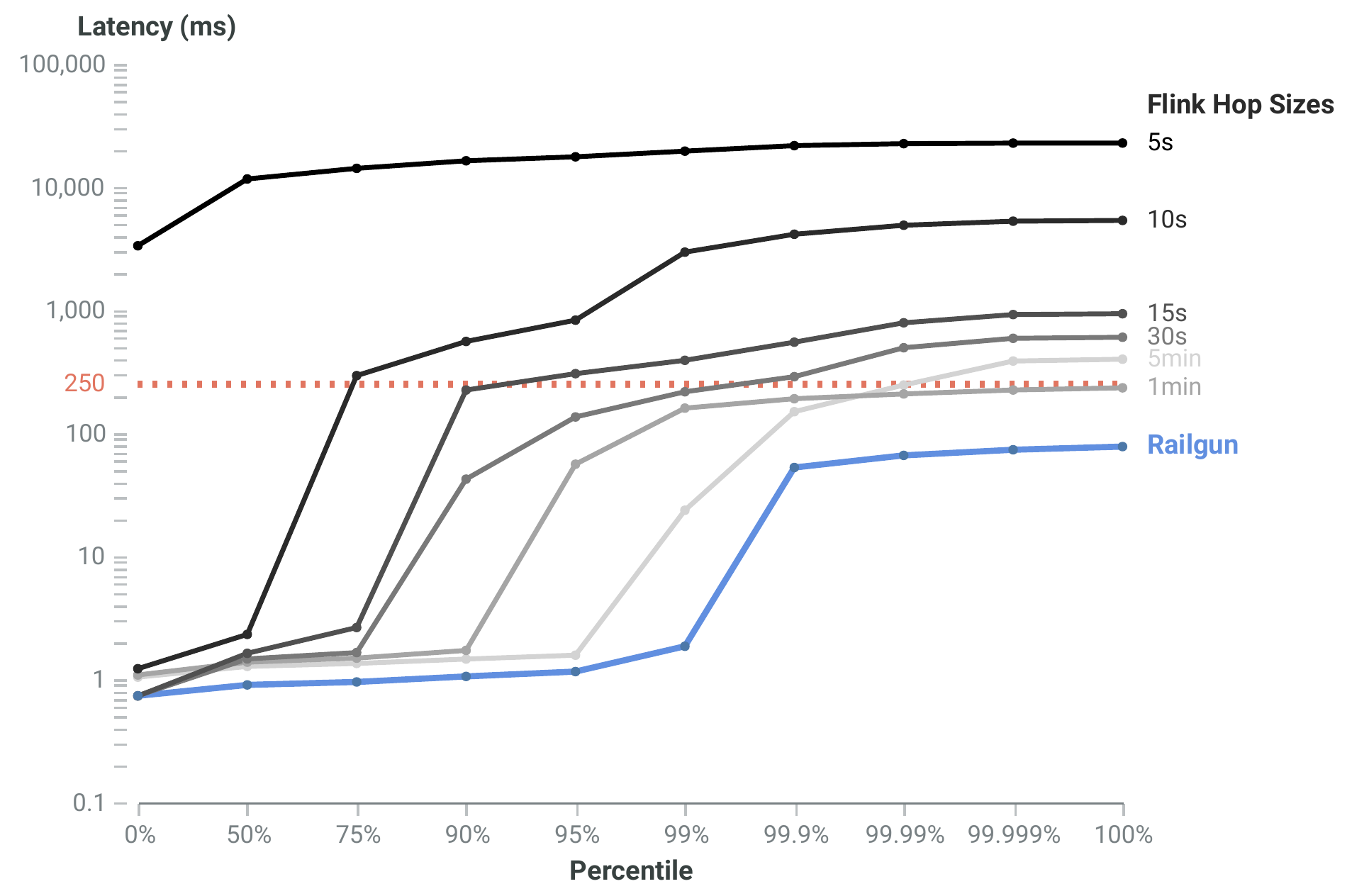}
  \caption{Distribution of Flink's latencies using hopping windows vs. Railgun's latencies using real-time sliding windows, at a fixed throughput of 500 ev/sec.}
  \label{fig:flink_vs_railgun}
\end{figure}

The results are shown in Figure~\ref{fig:flink_vs_railgun}, where we include Railgun's latency for the same query using its real-time sliding window.

\subsubsection{Discussion}
Figure~\ref{fig:flink_vs_railgun} shows how Flink latencies are affected when we increase the hop's granularity, i.e., when we increase the accuracy of the hopping window. Clearly, in this setup, with hops of 10s or less, Flink is unable to keep with a 500 ev/s throughput.
%
%
Furthermore, recall that in most of our setups, we are required to score events in less than 250ms in the 99.9\% percentile (cf. \textbf{M} requirement of Section~\ref{sec:realtimewindows}). For those, we need hops of at least 1 minute, which would severely compromise accuracy, and violate rules for our clients (cf. \textbf{A} requirement of Section~\ref{sec:realtimewindows}).
We would expect that with larger hops, e.g., at 10-min, or 30-min hops, Flink would have lower latencies than Railgun, but those hops would produce even bigger aggregation inaccuracies.
Railgun solves all these MAD requirements by using real-time sliding windows, accurate for every event, with lower latencies than Flink on all percentiles, on all windows using 1-min hops or less.

\subsection{Scaling Railgun Windows}
\label{subsec:scalingRailgun}
In our second set of experiments, we aim to demonstrate how Railgun performs when we scale the size of the window, and the number of windows, within a single machine. For this we designed two different experiments. For the first experiment \textbf{(a)}, we aim to show our claim that the window length is irrelevant for Railgun's performance. For that, we compute the same metric as in Section~\ref{subsection:flinkVsRailgun}, but vary the window size from 5 minutes to 7 days. Since our experiment runs are of 35 minutes (with 5 minute of warmup), and our largest window has 7 days, we start these experiments after a data checkpoint load, to ensure that windows are always iterating events for both its head and tail iterator.

For the second experiment \textbf{(b)}, we compute three different metrics: \texttt{sum, average} and \texttt{count}, over the \texttt{amount} field grouped by \texttt{card}. Then, we vary the number of windows on which we compute these three metrics, to enforce a different number of reservoir iterators. 
Recall from Section~\ref{subsec:reservoir} that a reservoir iterator is what hold event chunks in-memory for a given window. As depicted in Figure~\ref{fig:reservoir}, normally, each iterator will hold two chunks in-memory. The window's head iterator, receiving incoming events, holds the open chunk on which new events are being appended, and might still hold one\footnote{more closed chunks might still be in-memory, if the reservoir is configured for extensive support of out-of-order events, or if there is contention on the disk I/O.} closed chunk which is being written to disk asynchronously. Likewise, the window's tail iterator, iterating over the window's expiring events, necessarily holds the current chunk being iterated in-memory, and if possible, preemptively requests the following chunk to be loaded in the reservoir cache. Chunks might not be loaded if the reservoir cache is full. This might happen, if there are many iterators simultaneously reading reservoir chunks. Accessing a chunk not from cache can cause tail latency spikes. In the best scenario, these chunks are in the OS page cache and we only pay the cost for decompressing and deserializing. In the worst case, we also pay the full I/O seek cost.

The number of unique iterators depends on the number of windows configured in the system and how they are aligned. When two windows are aligned, either at the beginning or at the end, they share the same iterator. As such, for experiment (\textbf{b}), we force iterator misalignment by using windows with different window sizes and window delays. Namely, to vary from 20 to 240 iterators, in this experiment, we vary from 10 to 120 misaligned windows. 

\begin{figure}
    \centering
    \subfloat[Vary Window Size]{{\includegraphics[width=0.45\textwidth]{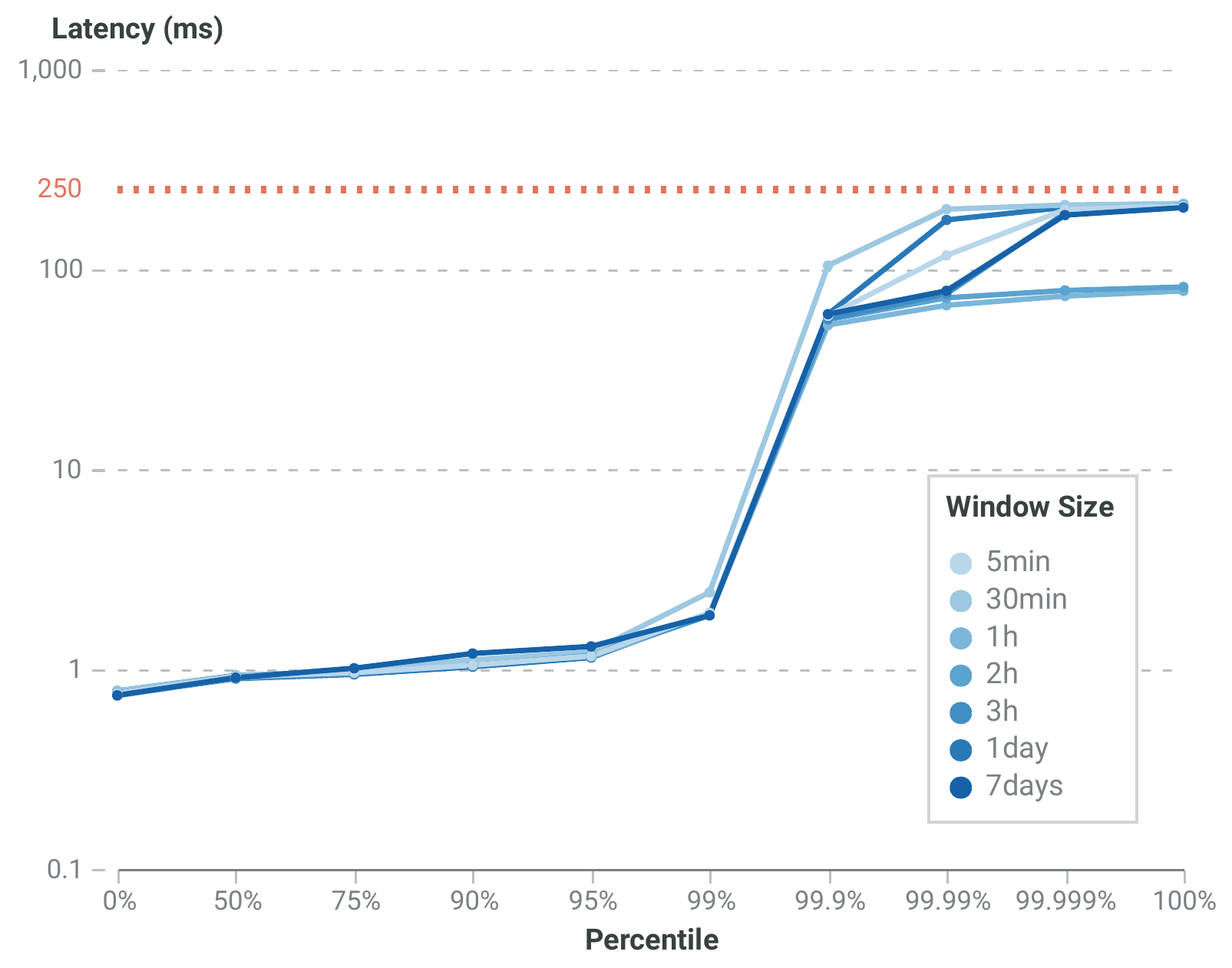} }}%
    \qquad
    \subfloat[Vary Number of Iterators]{{\includegraphics[width=0.45\textwidth]{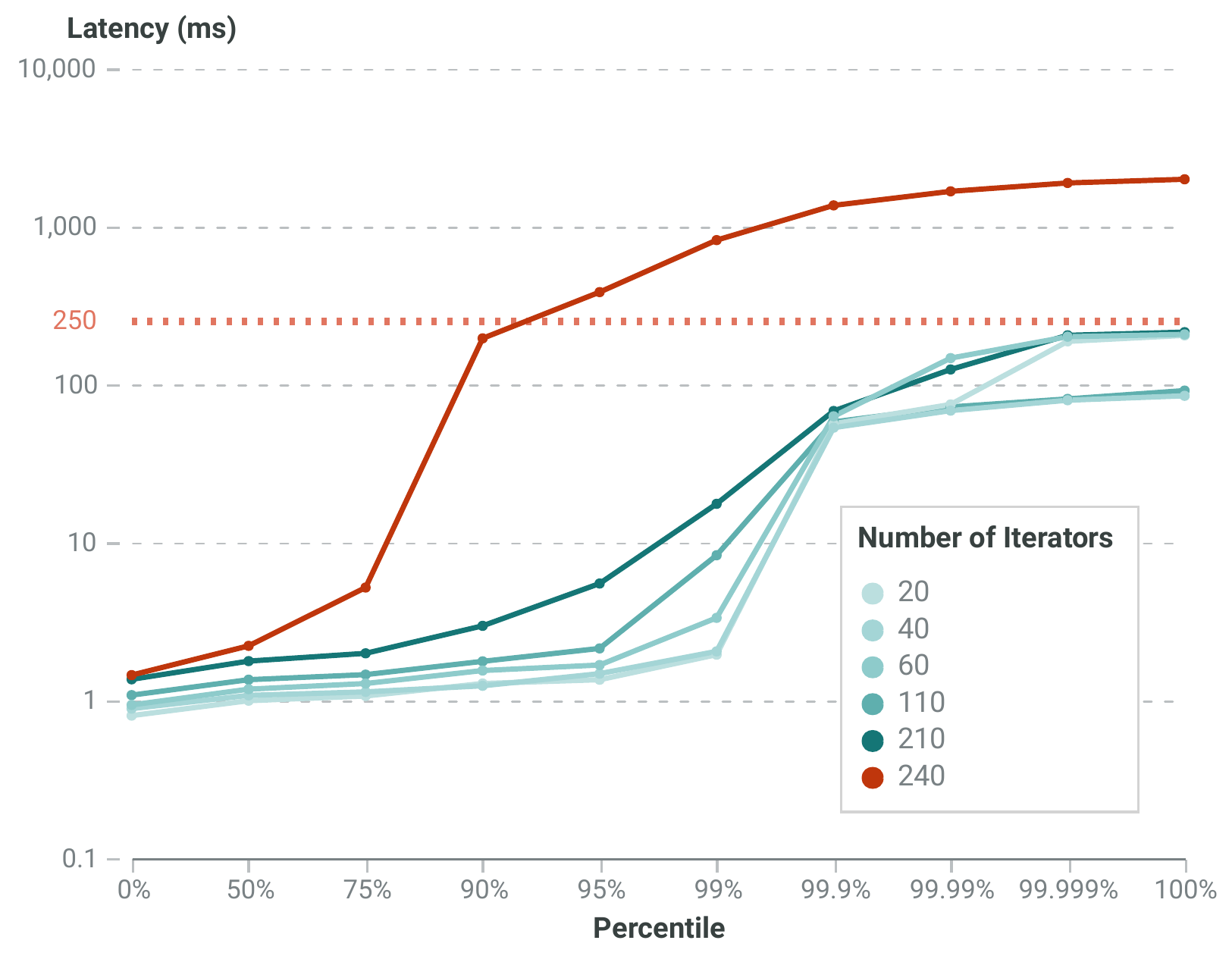} }}%
    \caption{Distribution of Railgun's latencies when scaling the window size and the number of windows, within a single machine, at a fixed throughput of 500 ev/s.}%
    \label{fig:scaling-railgun}%
\end{figure}

The results from both experiments can be seen in Figure~\ref{fig:scaling-railgun}.

\subsubsection{Discussion}
On experiment \textbf{(a)}, we clearly show that the window size is irrelevant to Railgun's latency performance. This is expected since for any window we have two iterators, independently of the window size. 
Additional benchmarks have shown us that variations in the higher percentiles (i.e., >99.9\%), are due to Kafka communication, rather than Railgun (and something that also affects Flink in Figure~\ref{fig:flink_vs_railgun}). Hence, in some runs we have ~150ms in 99.99\% percentile, while in others ~75ms. 

For experiment \textbf{(b)}, we show that as long as the iterators can retrieve the next chunk from cache, the impact on latencies is almost irrelevant. Each iterator requires a chunk in-memory, and, in this experiment, we used 220 chunk elements in Railgun's cache. This means that for most values used in this experiment (viz. between 20-210 iterators), whenever the iterator requests the next chunk, it is already available for iteration in the cache. Hence, we only start to see some latency degradation when we have almost the same elements in cache as the number of iterators, i.e., when we increase the probability of a cache-miss.
On the run where we have 240 iterators, we also start to see Garbage Collection (GC) problems due to memory pressure, which then leads to higher latencies. This is as expected since the actual heap usage is very close to the maximum JVM heap (10GB). 

\subsection{Scaling Railgun Nodes}
\label{subsec:distributedRailgun}
\begin{figure}[h]
  \includegraphics[width=0.45\textwidth]{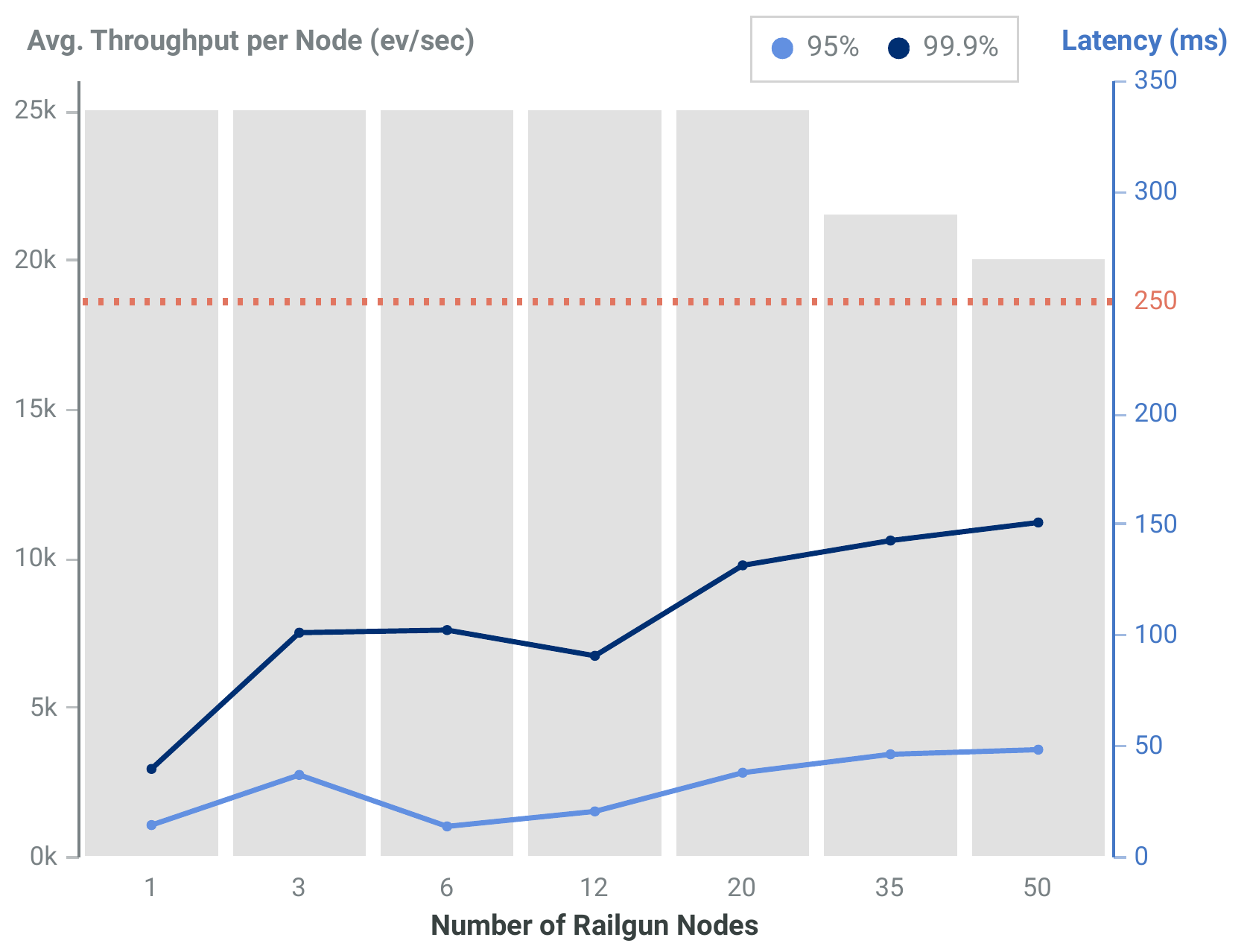}
  \caption{Evolution of the average throughput per node, when varying the throughput from 25 thousand ev/sec to 1 million ev/sec, and the number of nodes from 1 to 50.}
  \label{fig:distributedRailgun}
\end{figure}

In the last set of experiments, we aim to demonstrate how Railgun scales to address higher throughputs in multi-node setup, while still respecting our target Msec latency requirement at high percentiles (<250ms @ 99.9\%). Here we compute the same three metrics: \verb|sum|, \verb|average| and \verb|count| of \verb|amount| field grouped by \verb|card| over a 5-min window, and configured Railgun node to process as much load as possible, in a sustained way, without breaching the \textbf{M} requirement. For these machines (AWS \emph{m5.4xlarge} with 16 vCPUs and 64GB of RAM) we found the best performance using 8 Railgun processors per node and a JVM heap of 32GB (to take best advantage of compressed object pointers~\cite{compressedoops}),
where we could comfortably handle loads of 25 thousand ev/sec. 
Afterwards, we set our target to 1 million ev/sec, and increased our Railgun nodes gradually to achieve this target. 
The results are shown in Figure~\ref{fig:distributedRailgun}.

\subsubsection{Discussion}
On this experiment, we show that Railgun scales almost linearly, where we only start to see some small degradation with 35 nodes at a combined throughput of 750 thousand of ev/sec. Our target load of 1 million ev/sec can be achieved using a Railgun cluster with 50 nodes, where each node is processing, on average, 20 thousand ev/sec.

When analyzing this experiment results for a single node, we understood that our main bottleneck when handling these high loads is memory pressure and GC performance. Namely, at 25 thousand ev/sec, we are creating objects at a rate of about 5GB/sec. Although our heap of live objects occupies less than 7GB, at this creation rate, the GC struggles to keep up. To achieve higher throughputs per node we need to change our system to use off-heap memory management optimizations, as frequently done in other streaming engines~\cite{DBLP:journals/tocs/ShiKZJLZHHW19, flink-memory, spark-memory}.
Moreover, when scaling the cluster with multiple nodes, we start to see a bottleneck in Kafka, probably caused by the increased number of partitions needed to support the concurrent consumption of messages from the multiple nodes' processors. This is something to be improved in the future with a more careful tuning of Kafka configurations, and broker setup.


\section{Conclusions}
In this paper we propose Railgun, a novel distributed streaming engine that supports aggregations over real-time sliding windows, while providing crucial non-functional requirements: high throughput, tail low latency, horizontal scalability and fault tolerance.  

One of Railgun's most important enablers is the event reservoir. Since accurate metrics require considering all events, the reservoir efficiently persists them to disk, while fetching chunks of events ahead of time as they are needed by windows. This allows Railgun to support time-windows spanning years with the same memory usage as windows of seconds. 
To reduce storage costs, the reservoir uses cheap local HDDs or network-attached disks, and exploits the events' immutability to aggressively compress and serialize them.

Another central piece of Railgun's performance is how it takes advantage of Kafka to achieve a distributed, scalable and fault-tolerant system. In particular, we use the concept of \tp{} to delegate tasks among the several Railgun processors and exploit Kafka consumer's group guarantees to safeguard tasks assignment to a Railgun processors. Finally, to optimize recovery, we also provide a custom rebalance assignment strategy that minimizes data shuffle and maximizes load distribution across the cluster.

Railgun is still under development, and some work lies ahead to validate some components of our design. Namely, we need to certify that: 1) rebalance and recovery can be done respecting our latency requirements over the 99.9 percentile, especially when a task is assigned to a processor that has no previous data for that task; 2) we can efficiently support metrics backfill, i.e., the ability to add a new metric and fill it from old event data.

Although still a prototype, our experiments provide sound prom-ises for Railgun. Particularly, we show that Railgun has lower latencies per event than Flink, even when Flink is configured for a low metric accuracy (e.g., a 5-min hop size for a 60-min window). In addition, we show that our performance is unaffected by the window size, and that Railgun scales reasonably well with the number of windows and metrics, as we are able to prevent I/O calls on the critical path of event processing, by pro-actively loading reservoir chunks into memory, ahead of time.

Lastly, we demonstrate that Railgun can scale nearly linearly and process throughputs of million of events per second. Surely, there are many streaming engine systems achieving higher throughputs per node than Railgun, including Flink~\cite{DBLP:journals/debu/CarboneKEMHT15}, Spade~\cite{spade}, Kafka Streaming~\cite{sparkstreaming} or Spark Streaming~\cite{sparkstreaming}. However, to achieve it, these systems have to degrade latency at high percentiles, or make significant compromises on sliding window aggregation precision.
%
To the best of our knowledge, Railgun is the first distributed streaming engine able to deliver accurate real-time sliding window aggregations, with millisecond-level latencies at high percentiles, thereby making it the first of MAD systems.



\bibliographystyle{ACM-Reference-Format}
\bibliography{refs}


\begin{thebibliography}{53}


\ifx \showCODEN    \undefined \def \showCODEN     #1{\unskip}     \fi
\ifx \showDOI      \undefined \def \showDOI       #1{#1}\fi
\ifx \showISBNx    \undefined \def \showISBNx     #1{\unskip}     \fi
\ifx \showISBNxiii \undefined \def \showISBNxiii  #1{\unskip}     \fi
\ifx \showISSN     \undefined \def \showISSN      #1{\unskip}     \fi
\ifx \showLCCN     \undefined \def \showLCCN      #1{\unskip}     \fi
\ifx \shownote     \undefined \def \shownote      #1{#1}          \fi
\ifx \showarticletitle \undefined \def \showarticletitle #1{#1}   \fi
\ifx \showURL      \undefined \def \showURL       {\relax}        \fi
\providecommand\bibfield[2]{#2}
\providecommand\bibinfo[2]{#2}
\providecommand\natexlab[1]{#1}
\providecommand\showeprint[2][]{arXiv:#2}

\bibitem[\protect\citeauthoryear{Abadi, Ahmad, Balazinska, {\c{C}}., Cherniack,
  Hwang, Lindner, Maskey, Rasin, Ryvkina, Tatbul, Xing, and Zdonik}{Abadi
  et~al\mbox{.}}{2005a}]%
        {DBLP:conf/cidr/AbadiABCCHLMRRTXZ05}
\bibfield{author}{\bibinfo{person}{D.~J. Abadi}, \bibinfo{person}{Y. Ahmad},
  \bibinfo{person}{M. Balazinska}, \bibinfo{person}{U. {\c{C}}.},
  \bibinfo{person}{M. Cherniack}, \bibinfo{person}{J. Hwang},
  \bibinfo{person}{W. Lindner}, \bibinfo{person}{A. Maskey},
  \bibinfo{person}{A. Rasin}, \bibinfo{person}{E. Ryvkina}, \bibinfo{person}{N.
  Tatbul}, \bibinfo{person}{Y. Xing}, {and} \bibinfo{person}{S.~B. Zdonik}.}
  \bibinfo{year}{2005}\natexlab{a}.
\newblock \showarticletitle{The Design of the Borealis Stream Processing
  Engine}. In \bibinfo{booktitle}{\emph{{CIDR} 2005}}.
  \bibinfo{pages}{277--289}.
\newblock


\bibitem[\protect\citeauthoryear{Abadi, Ahmad, Balazinska, {\c{C}}etintemel,
  Cherniack, Hwang, Lindner, Maskey, Rasin, Ryvkina, Tatbul, Xing, and
  Zdonik}{Abadi et~al\mbox{.}}{2005b}]%
        {borealis}
\bibfield{author}{\bibinfo{person}{Daniel~J. Abadi}, \bibinfo{person}{Yanif
  Ahmad}, \bibinfo{person}{Magdalena Balazinska}, \bibinfo{person}{Ugur
  {\c{C}}etintemel}, \bibinfo{person}{Mitch Cherniack},
  \bibinfo{person}{Jeong{-}Hyon Hwang}, \bibinfo{person}{Wolfgang Lindner},
  \bibinfo{person}{Anurag Maskey}, \bibinfo{person}{Alex Rasin},
  \bibinfo{person}{Esther Ryvkina}, \bibinfo{person}{Nesime Tatbul},
  \bibinfo{person}{Ying Xing}, {and} \bibinfo{person}{Stanley~B. Zdonik}.}
  \bibinfo{year}{2005}\natexlab{b}.
\newblock \showarticletitle{The Design of the Borealis Stream Processing
  Engine}. In \bibinfo{booktitle}{\emph{{CIDR} 2005, Second Biennial Conference
  on Innovative Data Systems Research, Asilomar, CA, USA, January 4-7, 2005,
  Online Proceedings}}. \bibinfo{publisher}{www.cidrdb.org},
  \bibinfo{pages}{277--289}.
\newblock
\urldef\tempurl%
\url{http://cidrdb.org/cidr2005/papers/P23.pdf}
\showURL{%
\tempurl}


\bibitem[\protect\citeauthoryear{Abadi, Carney, {\c{C}}., Cherniack, Convey,
  Erwin, Galvez, Hatoun, Maskey, Rasin, Singer, Stonebraker, Tatbul, Xing, Yan,
  and Zdonik}{Abadi et~al\mbox{.}}{2003}]%
        {DBLP:conf/sigmod/AbadiCCCCEGHMRSSTXYZ03}
\bibfield{author}{\bibinfo{person}{D.~J. Abadi}, \bibinfo{person}{D. Carney},
  \bibinfo{person}{U. {\c{C}}.}, \bibinfo{person}{M. Cherniack},
  \bibinfo{person}{C. Convey}, \bibinfo{person}{C. Erwin},
  \bibinfo{person}{E.~F. Galvez}, \bibinfo{person}{M. Hatoun},
  \bibinfo{person}{A. Maskey}, \bibinfo{person}{A. Rasin}, \bibinfo{person}{A.
  Singer}, \bibinfo{person}{M. Stonebraker}, \bibinfo{person}{N. Tatbul},
  \bibinfo{person}{Y. Xing}, \bibinfo{person}{R. Yan}, {and}
  \bibinfo{person}{S.~B. Zdonik}.} \bibinfo{year}{2003}\natexlab{}.
\newblock \showarticletitle{Aurora: {A} Data Stream Management System}. In
  \bibinfo{booktitle}{\emph{{ACM} {SIGMOD} 2003}}. \bibinfo{publisher}{{ACM}},
  \bibinfo{pages}{666}.
\newblock
\urldef\tempurl%
\url{https://doi.org/10.1145/872757.872855}
\showDOI{\tempurl}


\bibitem[\protect\citeauthoryear{Akidau, Bradshaw, Chambers, Chernyak,
  Fern{\'{a}}ndez{-}Moctezuma, Lax, McVeety, Mills, Perry, Schmidt, and
  Whittle}{Akidau et~al\mbox{.}}{2015}]%
        {dataflow}
\bibfield{author}{\bibinfo{person}{Tyler Akidau}, \bibinfo{person}{Robert
  Bradshaw}, \bibinfo{person}{Craig Chambers}, \bibinfo{person}{Slava
  Chernyak}, \bibinfo{person}{Rafael Fern{\'{a}}ndez{-}Moctezuma},
  \bibinfo{person}{Reuven Lax}, \bibinfo{person}{Sam McVeety},
  \bibinfo{person}{Daniel Mills}, \bibinfo{person}{Frances Perry},
  \bibinfo{person}{Eric Schmidt}, {and} \bibinfo{person}{Sam Whittle}.}
  \bibinfo{year}{2015}\natexlab{}.
\newblock \showarticletitle{The Dataflow Model: {A} Practical Approach to
  Balancing Correctness, Latency, and Cost in Massive-Scale, Unbounded,
  Out-of-Order Data Processing}.
\newblock \bibinfo{journal}{\emph{Proc. {VLDB} Endow.}} \bibinfo{volume}{8},
  \bibinfo{number}{12} (\bibinfo{year}{2015}), \bibinfo{pages}{1792--1803}.
\newblock
\urldef\tempurl%
\url{https://doi.org/10.14778/2824032.2824076}
\showDOI{\tempurl}


\bibitem[\protect\citeauthoryear{Alibaba}{Alibaba}{2020}]%
        {alibaba-cloud}
\bibfield{author}{\bibinfo{person}{Alibaba}.} \bibinfo{year}{2020}\natexlab{}.
\newblock \bibinfo{title}{Alibaba Cloud Supported 583,000 Orders/Second for
  2020 Double 11}.
\newblock
  \bibinfo{howpublished}{\url{https://www.alibabacloud.com/blog/alibaba-cloud-supported-583000-orderssecond-for-2020-double-11---the-highest-traffic-peak-in-the-world_596884}}.
\newblock


\bibitem[\protect\citeauthoryear{Arasu, Babcock, Babu, Datar, Ito, Nishizawa,
  Rosenstein, and Widom}{Arasu et~al\mbox{.}}{2003}]%
        {DBLP:conf/sigmod/ArasuBBDIRW03}
\bibfield{author}{\bibinfo{person}{A. Arasu}, \bibinfo{person}{B. Babcock},
  \bibinfo{person}{S. Babu}, \bibinfo{person}{M. Datar}, \bibinfo{person}{K.
  Ito}, \bibinfo{person}{I. Nishizawa}, \bibinfo{person}{J. Rosenstein}, {and}
  \bibinfo{person}{J. Widom}.} \bibinfo{year}{2003}\natexlab{}.
\newblock \showarticletitle{{STREAM:} The Stanford Stream Data Manager}. In
  \bibinfo{booktitle}{\emph{{SIGMOD} 2003}}.
\newblock


\bibitem[\protect\citeauthoryear{Armbrust, Das, Torres, Yavuz, Zhu, Xin,
  Ghodsi, Stoica, and Zaharia}{Armbrust et~al\mbox{.}}{2018}]%
        {sparkstreaming}
\bibfield{author}{\bibinfo{person}{Michael Armbrust},
  \bibinfo{person}{Tathagata Das}, \bibinfo{person}{Joseph Torres},
  \bibinfo{person}{Burak Yavuz}, \bibinfo{person}{Shixiong Zhu},
  \bibinfo{person}{Reynold Xin}, \bibinfo{person}{Ali Ghodsi},
  \bibinfo{person}{Ion Stoica}, {and} \bibinfo{person}{Matei Zaharia}.}
  \bibinfo{year}{2018}\natexlab{}.
\newblock \showarticletitle{Structured Streaming: {A} Declarative {API} for
  Real-Time Applications in Apache Spark}. In
  \bibinfo{booktitle}{\emph{Proceedings of the 2018 International Conference on
  Management of Data, {SIGMOD} Conference 2018, Houston, TX, USA, June 10-15,
  2018}}, \bibfield{editor}{\bibinfo{person}{Gautam Das},
  \bibinfo{person}{Christopher~M. Jermaine}, {and} \bibinfo{person}{Philip~A.
  Bernstein}} (Eds.). \bibinfo{publisher}{{ACM}}, \bibinfo{pages}{601--613}.
\newblock
\urldef\tempurl%
\url{https://doi.org/10.1145/3183713.3190664}
\showDOI{\tempurl}


\bibitem[\protect\citeauthoryear{Babcock, Babu, Datar, Motwani, and
  Widom}{Babcock et~al\mbox{.}}{2002}]%
        {BabcockBDMW02}
\bibfield{author}{\bibinfo{person}{Brian Babcock}, \bibinfo{person}{Shivnath
  Babu}, \bibinfo{person}{Mayur Datar}, \bibinfo{person}{Rajeev Motwani}, {and}
  \bibinfo{person}{Jennifer Widom}.} \bibinfo{year}{2002}\natexlab{}.
\newblock \showarticletitle{Models and Issues in Data Stream Systems}. In
  \bibinfo{booktitle}{\emph{Proceedings of the Twenty-first {ACM}
  {SIGACT-SIGMOD-SIGART} Symposium on Principles of Database Systems, June 3-5,
  Madison, Wisconsin, {USA}}}, \bibfield{editor}{\bibinfo{person}{Lucian Popa},
  \bibinfo{person}{Serge Abiteboul}, {and} \bibinfo{person}{Phokion~G.
  Kolaitis}} (Eds.). \bibinfo{publisher}{{ACM}}, \bibinfo{pages}{1--16}.
\newblock
\urldef\tempurl%
\url{https://doi.org/10.1145/543613.543615}
\showDOI{\tempurl}


\bibitem[\protect\citeauthoryear{Benson, Grulich, Zeuch, Markl, and
  Rabl}{Benson et~al\mbox{.}}{2020}]%
        {DBLP:conf/edbt/BensonGZMR20}
\bibfield{author}{\bibinfo{person}{L. Benson}, \bibinfo{person}{P.~M. Grulich},
  \bibinfo{person}{S. Zeuch}, \bibinfo{person}{V. Markl}, {and}
  \bibinfo{person}{T. Rabl}.} \bibinfo{year}{2020}\natexlab{}.
\newblock \showarticletitle{Disco: Efficient Distributed Window Aggregation}.
  In \bibinfo{booktitle}{\emph{{EDBT}'20}}.
  \bibinfo{publisher}{OpenProceedings.org}, \bibinfo{pages}{423--426}.
\newblock


\bibitem[\protect\citeauthoryear{Branco, Abreu, Gomes, Almeida, Ascens{\~{a}}o,
  and Bizarro}{Branco et~al\mbox{.}}{2020}]%
        {kdd-Branco-deeplearning}
\bibfield{author}{\bibinfo{person}{Bernardo Branco}, \bibinfo{person}{Pedro
  Abreu}, \bibinfo{person}{Ana~Sofia Gomes}, \bibinfo{person}{Mariana S.~C.
  Almeida}, \bibinfo{person}{Jo{\~{a}}o~Tiago Ascens{\~{a}}o}, {and}
  \bibinfo{person}{Pedro Bizarro}.} \bibinfo{year}{2020}\natexlab{}.
\newblock \showarticletitle{Interleaved Sequence RNNs for Fraud Detection}. In
  \bibinfo{booktitle}{\emph{{KDD} '20: The 26th {ACM} {SIGKDD} Conference on
  Knowledge Discovery and Data Mining, Virtual Event, CA, USA, August 23-27,
  2020}}, \bibfield{editor}{\bibinfo{person}{Rajesh Gupta},
  \bibinfo{person}{Yan Liu}, \bibinfo{person}{Jiliang Tang}, {and}
  \bibinfo{person}{B.~Aditya Prakash}} (Eds.). \bibinfo{publisher}{{ACM}},
  \bibinfo{pages}{3101--3109}.
\newblock
\showISBNx{978-1-4503-7998-4}
\urldef\tempurl%
\url{https://doi.org/10.1145/3394486}
\showDOI{\tempurl}


\bibitem[\protect\citeauthoryear{Carbone, Katsifodimos, Ewen, Markl, Haridi,
  and Tzoumas}{Carbone et~al\mbox{.}}{2015}]%
        {DBLP:journals/debu/CarboneKEMHT15}
\bibfield{author}{\bibinfo{person}{Paris Carbone}, \bibinfo{person}{Asterios
  Katsifodimos}, \bibinfo{person}{Stephan Ewen}, \bibinfo{person}{Volker
  Markl}, \bibinfo{person}{Seif Haridi}, {and} \bibinfo{person}{Kostas
  Tzoumas}.} \bibinfo{year}{2015}\natexlab{}.
\newblock \showarticletitle{Apache Flink{\texttrademark}: Stream and Batch
  Processing in a Single Engine}.
\newblock \bibinfo{journal}{\emph{{IEEE} Data Eng. Bull.}}
  \bibinfo{volume}{38}, \bibinfo{number}{4} (\bibinfo{year}{2015}),
  \bibinfo{pages}{28--38}.
\newblock
\urldef\tempurl%
\url{http://sites.computer.org/debull/A15dec/p28.pdf}
\showURL{%
\tempurl}


\bibitem[\protect\citeauthoryear{Carbone, Traub, Katsifodimos, Haridi, and
  Markl}{Carbone et~al\mbox{.}}{2016}]%
        {DBLP:conf/cikm/CarboneTKHM16}
\bibfield{author}{\bibinfo{person}{P. Carbone}, \bibinfo{person}{J. Traub},
  \bibinfo{person}{A. Katsifodimos}, \bibinfo{person}{S. Haridi}, {and}
  \bibinfo{person}{V. Markl}.} \bibinfo{year}{2016}\natexlab{}.
\newblock \showarticletitle{Cutty: Aggregate Sharing for User-Defined Windows}.
  In \bibinfo{booktitle}{\emph{{CIKM}'16}}. \bibinfo{publisher}{{ACM}}.
\newblock


\bibitem[\protect\citeauthoryear{Chandramouli, Goldstein, Barnett, and
  Terwilliger}{Chandramouli et~al\mbox{.}}{2015}]%
        {trill}
\bibfield{author}{\bibinfo{person}{Badrish Chandramouli},
  \bibinfo{person}{Jonathan Goldstein}, \bibinfo{person}{Mike Barnett}, {and}
  \bibinfo{person}{James~F. Terwilliger}.} \bibinfo{year}{2015}\natexlab{}.
\newblock \showarticletitle{Trill: Engineering a Library for Diverse
  Analytics}.
\newblock \bibinfo{journal}{\emph{{IEEE} Data Eng. Bull.}}
  \bibinfo{volume}{38}, \bibinfo{number}{4} (\bibinfo{year}{2015}),
  \bibinfo{pages}{51--60}.
\newblock
\urldef\tempurl%
\url{http://sites.computer.org/debull/A15dec/p51.pdf}
\showURL{%
\tempurl}


\bibitem[\protect\citeauthoryear{Chandrasekaran, Cooper, Deshpande, Franklin,
  Hellerstein, Hong, Krishnamurthy, Madden, Raman, Reiss, and
  Shah}{Chandrasekaran et~al\mbox{.}}{2003}]%
        {DBLP:conf/cidr/ChandrasekaranDFHHKMRRS03}
\bibfield{author}{\bibinfo{person}{S. Chandrasekaran}, \bibinfo{person}{O.
  Cooper}, \bibinfo{person}{A. Deshpande}, \bibinfo{person}{M.~J. Franklin},
  \bibinfo{person}{J.~M. Hellerstein}, \bibinfo{person}{W. Hong},
  \bibinfo{person}{S. Krishnamurthy}, \bibinfo{person}{S. Madden},
  \bibinfo{person}{V. Raman}, \bibinfo{person}{F. Reiss}, {and}
  \bibinfo{person}{M.~A. Shah}.} \bibinfo{year}{2003}\natexlab{}.
\newblock \showarticletitle{TelegraphCQ: Continuous Dataflow Processing for
  Uncertain World}. In \bibinfo{booktitle}{\emph{{CIDR}'03}}.
\newblock


\bibitem[\protect\citeauthoryear{Chatterjee and Morin}{Chatterjee and
  Morin}{2018}]%
        {ChatterjeeM18}
\bibfield{author}{\bibinfo{person}{Subarna Chatterjee} {and}
  \bibinfo{person}{Christine Morin}.} \bibinfo{year}{2018}\natexlab{}.
\newblock \showarticletitle{Experimental Study on the Performance and Resource
  Utilization of Data Streaming Frameworks}. In \bibinfo{booktitle}{\emph{18th
  {IEEE/ACM} International Symposium on Cluster, Cloud and Grid Computing,
  {CCGRID} 2018, Washington, DC, USA, May 1-4, 2018}},
  \bibfield{editor}{\bibinfo{person}{Esam El{-}Araby},
  \bibinfo{person}{Dhabaleswar~K. Panda}, \bibinfo{person}{Sandra Gesing},
  \bibinfo{person}{Amy~W. Apon}, \bibinfo{person}{Volodymyr~V. Kindratenko},
  \bibinfo{person}{Massimo Cafaro}, {and} \bibinfo{person}{Alfredo Cuzzocrea}}
  (Eds.). \bibinfo{publisher}{{IEEE} Computer Society},
  \bibinfo{pages}{143--152}.
\newblock
\urldef\tempurl%
\url{https://doi.org/10.1109/CCGRID.2018.00029}
\showDOI{\tempurl}


\bibitem[\protect\citeauthoryear{Chen, DeWitt, Tian, and Wang}{Chen
  et~al\mbox{.}}{2000}]%
        {niagaracq}
\bibfield{author}{\bibinfo{person}{Jianjun Chen}, \bibinfo{person}{David~J.
  DeWitt}, \bibinfo{person}{Feng Tian}, {and} \bibinfo{person}{Yuan Wang}.}
  \bibinfo{year}{2000}\natexlab{}.
\newblock \showarticletitle{NiagaraCQ: {A} Scalable Continuous Query System for
  Internet Databases}. In \bibinfo{booktitle}{\emph{Proceedings of the 2000
  {ACM} {SIGMOD} International Conference on Management of Data, May 16-18,
  2000, Dallas, Texas, {USA}}}, \bibfield{editor}{\bibinfo{person}{Weidong
  Chen}, \bibinfo{person}{Jeffrey~F. Naughton}, {and}
  \bibinfo{person}{Philip~A. Bernstein}} (Eds.). \bibinfo{publisher}{{ACM}},
  \bibinfo{pages}{379--390}.
\newblock
\showISBNx{1-58113-217-4}
\urldef\tempurl%
\url{https://doi.org/10.1145/342009.335432}
\showDOI{\tempurl}


\bibitem[\protect\citeauthoryear{Commons}{Commons}{2011}]%
        {jexel}
\bibfield{author}{\bibinfo{person}{Apache Commons}.}
  \bibinfo{year}{2011}\natexlab{}.
\newblock \bibinfo{title}{{JEXEL} Expressions}.
\newblock
  \bibinfo{howpublished}{\url{https://commons.apache.org/proper/commons-jexl/}}.
\newblock


\bibitem[\protect\citeauthoryear{Confluent}{Confluent}{2016}]%
        {kreps2016introducing}
\bibfield{author}{\bibinfo{person}{Confluent}.}
  \bibinfo{year}{2016}\natexlab{}.
\newblock \bibinfo{title}{Kafka Streams}.
\newblock
  \bibinfo{howpublished}{\url{https://www.confluent.io/blog/introducing-kafka-streams-stream-processing-made-simple/}}.
\newblock


\bibitem[\protect\citeauthoryear{Dean and Barroso}{Dean and Barroso}{2013}]%
        {DBLP:journals/cacm/DeanB13}
\bibfield{author}{\bibinfo{person}{Jeffrey Dean} {and}
  \bibinfo{person}{Luiz~Andr{\'{e}} Barroso}.} \bibinfo{year}{2013}\natexlab{}.
\newblock \showarticletitle{The tail at scale}.
\newblock \bibinfo{journal}{\emph{Commun. {ACM}}} \bibinfo{volume}{56},
  \bibinfo{number}{2} (\bibinfo{year}{2013}), \bibinfo{pages}{74--80}.
\newblock
\urldef\tempurl%
\url{https://doi.org/10.1145/2408776.2408794}
\showDOI{\tempurl}


\bibitem[\protect\citeauthoryear{Facebook}{Facebook}{2012}]%
        {rocksdb}
\bibfield{author}{\bibinfo{person}{Facebook}.} \bibinfo{year}{2012}\natexlab{}.
\newblock \bibinfo{title}{RocksDB}.
\newblock \bibinfo{howpublished}{\url{https://rocksdb.org/}}.
\newblock


\bibitem[\protect\citeauthoryear{Fedulov}{Fedulov}{2020}]%
        {flink-fraud}
\bibfield{author}{\bibinfo{person}{A. Fedulov}.}
  \bibinfo{year}{2020}\natexlab{}.
\newblock \bibinfo{title}{Advanced Flink Appl. Patterns Vol.3: Custom Window
  Processing}.
\newblock
  \bibinfo{howpublished}{\url{https://flink.apache.org/news/2020/07/30/demo-fraud-detection-3.html}}.
\newblock


\bibitem[\protect\citeauthoryear{Franklin, Krishnamurthy, Conway, Li,
  Russakovsky, and Thombre}{Franklin et~al\mbox{.}}{2009}]%
        {FranklinKCLRT09}
\bibfield{author}{\bibinfo{person}{Michael~J. Franklin},
  \bibinfo{person}{Sailesh Krishnamurthy}, \bibinfo{person}{Neil Conway},
  \bibinfo{person}{Alan Li}, \bibinfo{person}{Alex Russakovsky}, {and}
  \bibinfo{person}{Neil Thombre}.} \bibinfo{year}{2009}\natexlab{}.
\newblock \showarticletitle{Continuous Analytics: Rethinking Query Processing
  in a Network-Effect World}. In \bibinfo{booktitle}{\emph{Fourth Biennial
  Conference on Innovative Data Systems Research, {CIDR} 2009, Asilomar, CA,
  USA, January 4-7, 2009, Online Proceedings}}.
  \bibinfo{publisher}{www.cidrdb.org}.
\newblock
\urldef\tempurl%
\url{http://www-db.cs.wisc.edu/cidr/cidr2009/Paper\_122.pdf}
\showURL{%
\tempurl}


\bibitem[\protect\citeauthoryear{Gedik, Andrade, Wu, Yu, and Doo}{Gedik
  et~al\mbox{.}}{2008}]%
        {spade}
\bibfield{author}{\bibinfo{person}{Bugra Gedik}, \bibinfo{person}{Henrique
  Andrade}, \bibinfo{person}{Kun{-}Lung Wu}, \bibinfo{person}{Philip~S. Yu},
  {and} \bibinfo{person}{Myungcheol Doo}.} \bibinfo{year}{2008}\natexlab{}.
\newblock \showarticletitle{{SPADE:} the system s declarative stream processing
  engine}. In \bibinfo{booktitle}{\emph{Proceedings of the {ACM} {SIGMOD}
  International Conference on Management of Data, {SIGMOD} 2008, Vancouver, BC,
  Canada, June 10-12, 2008}},
  \bibfield{editor}{\bibinfo{person}{Jason~Tsong{-}Li Wang}} (Ed.).
  \bibinfo{publisher}{{ACM}}, \bibinfo{pages}{1123--1134}.
\newblock
\showISBNx{978-1-60558-102-6}
\urldef\tempurl%
\url{https://doi.org/10.1145/1376616.1376729}
\showDOI{\tempurl}


\bibitem[\protect\citeauthoryear{Golab and {\"{O}}zsu}{Golab and
  {\"{O}}zsu}{2003}]%
        {GolabO03}
\bibfield{author}{\bibinfo{person}{Lukasz Golab} {and}
  \bibinfo{person}{M.~Tamer {\"{O}}zsu}.} \bibinfo{year}{2003}\natexlab{}.
\newblock \showarticletitle{Issues in data stream management}.
\newblock \bibinfo{journal}{\emph{{SIGMOD} Rec.}} \bibinfo{volume}{32},
  \bibinfo{number}{2} (\bibinfo{year}{2003}), \bibinfo{pages}{5--14}.
\newblock
\urldef\tempurl%
\url{https://doi.org/10.1145/776985.776986}
\showDOI{\tempurl}


\bibitem[\protect\citeauthoryear{Hirzel, Soul{\'{e}}, Schneider, Gedik, and
  Grimm}{Hirzel et~al\mbox{.}}{2013}]%
        {HirzelSSGG13}
\bibfield{author}{\bibinfo{person}{Martin Hirzel}, \bibinfo{person}{Robert
  Soul{\'{e}}}, \bibinfo{person}{Scott Schneider}, \bibinfo{person}{Bugra
  Gedik}, {and} \bibinfo{person}{Robert Grimm}.}
  \bibinfo{year}{2013}\natexlab{}.
\newblock \showarticletitle{A catalog of stream processing optimizations}.
\newblock \bibinfo{journal}{\emph{{ACM} Comput. Surv.}} \bibinfo{volume}{46},
  \bibinfo{number}{4} (\bibinfo{year}{2013}), \bibinfo{pages}{46:1--46:34}.
\newblock
\urldef\tempurl%
\url{https://doi.org/10.1145/2528412}
\showDOI{\tempurl}


\bibitem[\protect\citeauthoryear{Hoff}{Hoff}{2015}]%
        {CoordOmi}
\bibfield{author}{\bibinfo{person}{T. Hoff}.} \bibinfo{year}{2015}\natexlab{}.
\newblock \bibinfo{title}{Your Load Generator Is Probably Lying To You - Take
  The Red Pill And Find Out Why}.
\newblock
  \bibinfo{howpublished}{\url{http://highscalability.com/blog/2015/10/5/your-load-generator-is-probably-lying-to-you-take-the-red-pi.html}}.
\newblock


\bibitem[\protect\citeauthoryear{Karimov, Rabl, Katsifodimos, Samarev,
  Heiskanen, and Markl}{Karimov et~al\mbox{.}}{2018}]%
        {KarimovRKSHM18}
\bibfield{author}{\bibinfo{person}{Jeyhun Karimov}, \bibinfo{person}{Tilmann
  Rabl}, \bibinfo{person}{Asterios Katsifodimos}, \bibinfo{person}{Roman
  Samarev}, \bibinfo{person}{Henri Heiskanen}, {and} \bibinfo{person}{Volker
  Markl}.} \bibinfo{year}{2018}\natexlab{}.
\newblock \showarticletitle{Benchmarking Distributed Stream Data Processing
  Systems}. In \bibinfo{booktitle}{\emph{34th {IEEE} International Conference
  on Data Engineering, {ICDE} 2018, Paris, France, April 16-19, 2018}}.
  \bibinfo{publisher}{{IEEE} Computer Society}, \bibinfo{pages}{1507--1518}.
\newblock
\urldef\tempurl%
\url{https://doi.org/10.1109/ICDE.2018.00169}
\showDOI{\tempurl}


\bibitem[\protect\citeauthoryear{Karp}{Karp}{1972}]%
        {DBLP:conf/coco/Karp72}
\bibfield{author}{\bibinfo{person}{Richard~M. Karp}.}
  \bibinfo{year}{1972}\natexlab{}.
\newblock \showarticletitle{Reducibility Among Combinatorial Problems}. In
  \bibinfo{booktitle}{\emph{Proceedings of a symposium on the Complexity of
  Computer Computations, held March 20-22, 1972, at the {IBM} Thomas J. Watson
  Research Center, Yorktown Heights, New York, {USA}}}
  \emph{(\bibinfo{series}{The {IBM} Research Symposia Series})},
  \bibfield{editor}{\bibinfo{person}{Raymond~E. Miller} {and}
  \bibinfo{person}{James~W. Thatcher}} (Eds.). \bibinfo{publisher}{Plenum
  Press, New York}, \bibinfo{pages}{85--103}.
\newblock
\showISBNx{0-306-30707-3}
\urldef\tempurl%
\url{https://doi.org/10.1007/978-1-4684-2001-2\_9}
\showDOI{\tempurl}


\bibitem[\protect\citeauthoryear{Kiran, Murphy, Monga, Dugan, and Baveja}{Kiran
  et~al\mbox{.}}{2015}]%
        {KiranMMDB15}
\bibfield{author}{\bibinfo{person}{Mariam Kiran}, \bibinfo{person}{Peter
  Murphy}, \bibinfo{person}{Inder Monga}, \bibinfo{person}{Jon Dugan}, {and}
  \bibinfo{person}{Sartaj~Singh Baveja}.} \bibinfo{year}{2015}\natexlab{}.
\newblock \showarticletitle{Lambda architecture for cost-effective batch and
  speed big data processing}. In \bibinfo{booktitle}{\emph{2015 {IEEE}
  International Conference on Big Data, Big Data 2015, Santa Clara, CA, USA,
  October 29 - November 1, 2015}}. \bibinfo{publisher}{{IEEE} Computer
  Society}, \bibinfo{pages}{2785--2792}.
\newblock
\urldef\tempurl%
\url{https://doi.org/10.1109/BigData.2015.7364082}
\showDOI{\tempurl}


\bibitem[\protect\citeauthoryear{Knuth}{Knuth}{1997}]%
        {donaldknuth}
\bibfield{author}{\bibinfo{person}{Donald~E. Knuth}.}
  \bibinfo{year}{1997}\natexlab{}.
\newblock \bibinfo{booktitle}{\emph{The Art of Computer Programming, Volume 1
  (3rd Ed.): Fundamental Algorithms}}.
\newblock \bibinfo{publisher}{Addison Wesley Longman Publishing Co., Inc.},
  \bibinfo{address}{USA}.
\newblock
\showISBNx{0201896834}


\bibitem[\protect\citeauthoryear{Kreps, Narkhede, Rao, et~al\mbox{.}}{Kreps
  et~al\mbox{.}}{2011}]%
        {kreps2011kafka}
\bibfield{author}{\bibinfo{person}{Jay Kreps}, \bibinfo{person}{Neha Narkhede},
  \bibinfo{person}{Jun Rao}, {et~al\mbox{.}}} \bibinfo{year}{2011}\natexlab{}.
\newblock \showarticletitle{Kafka: A distributed messaging system for log
  processing}. In \bibinfo{booktitle}{\emph{Proceedings of the NetDB}},
  Vol.~\bibinfo{volume}{11}. \bibinfo{pages}{1--7}.
\newblock


\bibitem[\protect\citeauthoryear{Kulkarni, Bhagat, Fu, Kedigehalli, Kellogg,
  Mittal, Patel, Ramasamy, and Taneja}{Kulkarni et~al\mbox{.}}{2015}]%
        {heronSIGMOD15}
\bibfield{author}{\bibinfo{person}{Sanjeev Kulkarni}, \bibinfo{person}{Nikunj
  Bhagat}, \bibinfo{person}{Maosong Fu}, \bibinfo{person}{Vikas Kedigehalli},
  \bibinfo{person}{Christopher Kellogg}, \bibinfo{person}{Sailesh Mittal},
  \bibinfo{person}{Jignesh~M. Patel}, \bibinfo{person}{Karthik Ramasamy}, {and}
  \bibinfo{person}{Siddarth Taneja}.} \bibinfo{year}{2015}\natexlab{}.
\newblock \showarticletitle{Twitter Heron: Stream Processing at Scale}. In
  \bibinfo{booktitle}{\emph{Proceedings of the 2015 {ACM} {SIGMOD}
  International Conference on Management of Data, Melbourne, Victoria,
  Australia, May 31 - June 4, 2015}},
  \bibfield{editor}{\bibinfo{person}{Timos~K. Sellis},
  \bibinfo{person}{Susan~B. Davidson}, {and} \bibinfo{person}{Zachary~G. Ives}}
  (Eds.). \bibinfo{publisher}{{ACM}}, \bibinfo{pages}{239--250}.
\newblock
\urldef\tempurl%
\url{https://doi.org/10.1145/2723372.2742788}
\showDOI{\tempurl}


\bibitem[\protect\citeauthoryear{Lightbend}{Lightbend}{2011}]%
        {Akka}
\bibfield{author}{\bibinfo{person}{Lightbend}.}
  \bibinfo{year}{2011}\natexlab{}.
\newblock \bibinfo{title}{Akka Streams}.
\newblock
  \bibinfo{howpublished}{\url{https://doc.akka.io/docs/akka/current/stream/index.html}}.
\newblock


\bibitem[\protect\citeauthoryear{Mendes, Bizarro, and Marques}{Mendes
  et~al\mbox{.}}{2013}]%
        {slidem}
\bibfield{author}{\bibinfo{person}{Marcelo R.~N. Mendes},
  \bibinfo{person}{Pedro Bizarro}, {and} \bibinfo{person}{Paulo Marques}.}
  \bibinfo{year}{2013}\natexlab{}.
\newblock \showarticletitle{Overcoming memory limitations in high-throughput
  event-based applications}. In \bibinfo{booktitle}{\emph{{ACM/SPEC}
  International Conference on Performance Engineering, ICPE'13, Prague, Czech
  Republic - April 21 - 24, 2013}},
  \bibfield{editor}{\bibinfo{person}{Seetharami Seelam}, \bibinfo{person}{Petr
  Tuma}, \bibinfo{person}{Giuliano Casale}, \bibinfo{person}{Tony Field}, {and}
  \bibinfo{person}{Jos{\'{e}}~Nelson Amaral}} (Eds.).
  \bibinfo{publisher}{{ACM}}, \bibinfo{pages}{399--410}.
\newblock
\showISBNx{978-1-4503-1636-1}
\urldef\tempurl%
\url{https://doi.org/10.1145/2479871.2479933}
\showDOI{\tempurl}


\bibitem[\protect\citeauthoryear{Navas}{Navas}{2010}]%
        {Navas10}
\bibfield{author}{\bibinfo{person}{Julio~J. Navas}.}
  \bibinfo{year}{2010}\natexlab{}.
\newblock \showarticletitle{Insight into Events: Event and Data Management for
  the Extended Enterprise}. In \bibinfo{booktitle}{\emph{Enabling Real-Time
  Business Intelligence - 4th International Workshop, {BIRTE} 2010, Held at the
  36th International Conference on Very Large Databases, {VLDB} 2010,
  Singapore, September 13, 2010, Revised Selected Papers}}
  \emph{(\bibinfo{series}{Lecture Notes in Business Information Processing})},
  \bibfield{editor}{\bibinfo{person}{Mal{\'{u}} Castellanos},
  \bibinfo{person}{Umeshwar Dayal}, {and} \bibinfo{person}{Volker Markl}}
  (Eds.), Vol.~\bibinfo{volume}{84}. \bibinfo{publisher}{Springer},
  \bibinfo{pages}{24--35}.
\newblock
\urldef\tempurl%
\url{https://doi.org/10.1007/978-3-642-22970-1\_3}
\showDOI{\tempurl}


\bibitem[\protect\citeauthoryear{Noghabi, Paramasivam, Pan, Ramesh, Bringhurst,
  Gupta, and Campbell}{Noghabi et~al\mbox{.}}{2017a}]%
        {noghabi2017samza}
\bibfield{author}{\bibinfo{person}{Shadi~A. Noghabi}, \bibinfo{person}{Kartik
  Paramasivam}, \bibinfo{person}{Yi Pan}, \bibinfo{person}{Navina Ramesh},
  \bibinfo{person}{Jon Bringhurst}, \bibinfo{person}{Indranil Gupta}, {and}
  \bibinfo{person}{Roy~H. Campbell}.} \bibinfo{year}{2017}\natexlab{a}.
\newblock \showarticletitle{Stateful Scalable Stream Processing at LinkedIn}.
\newblock \bibinfo{journal}{\emph{Proc. {VLDB} Endow.}} \bibinfo{volume}{10},
  \bibinfo{number}{12} (\bibinfo{year}{2017}), \bibinfo{pages}{1634--1645}.
\newblock
\urldef\tempurl%
\url{https://doi.org/10.14778/3137765.3137770}
\showDOI{\tempurl}


\bibitem[\protect\citeauthoryear{Noghabi, Paramasivam, Pan, Ramesh, Bringhurst,
  Gupta, and Campbell}{Noghabi et~al\mbox{.}}{2017b}]%
        {samzaVLDB17}
\bibfield{author}{\bibinfo{person}{Shadi~A. Noghabi}, \bibinfo{person}{Kartik
  Paramasivam}, \bibinfo{person}{Yi Pan}, \bibinfo{person}{Navina Ramesh},
  \bibinfo{person}{Jon Bringhurst}, \bibinfo{person}{Indranil Gupta}, {and}
  \bibinfo{person}{Roy~H. Campbell}.} \bibinfo{year}{2017}\natexlab{b}.
\newblock \showarticletitle{Stateful Scalable Stream Processing at LinkedIn}.
\newblock \bibinfo{journal}{\emph{Proc. {VLDB} Endow.}} \bibinfo{volume}{10},
  \bibinfo{number}{12} (\bibinfo{year}{2017}), \bibinfo{pages}{1634--1645}.
\newblock
\urldef\tempurl%
\url{https://doi.org/10.14778/3137765.3137770}
\showDOI{\tempurl}


\bibitem[\protect\citeauthoryear{O'Neil, Cheng, Gawlick, and O'Neil}{O'Neil
  et~al\mbox{.}}{1996}]%
        {ONeilCGO96}
\bibfield{author}{\bibinfo{person}{Patrick~E. O'Neil}, \bibinfo{person}{Edward
  Cheng}, \bibinfo{person}{Dieter Gawlick}, {and} \bibinfo{person}{Elizabeth~J.
  O'Neil}.} \bibinfo{year}{1996}\natexlab{}.
\newblock \showarticletitle{The Log-Structured Merge-Tree (LSM-Tree)}.
\newblock \bibinfo{journal}{\emph{Acta Inf.}} \bibinfo{volume}{33},
  \bibinfo{number}{4} (\bibinfo{year}{1996}), \bibinfo{pages}{351--385}.
\newblock
\urldef\tempurl%
\url{https://doi.org/10.1007/s002360050048}
\showDOI{\tempurl}


\bibitem[\protect\citeauthoryear{Ramasamy}{Ramasamy}{2019}]%
        {ramasamy2019unifying}
\bibfield{author}{\bibinfo{person}{Karthik Ramasamy}.}
  \bibinfo{year}{2019}\natexlab{}.
\newblock \showarticletitle{Unifying Messaging, Queuing, Streaming and Light
  Weight Compute for Online Event Processing}. In
  \bibinfo{booktitle}{\emph{Proceedings of the 13th {ACM} International
  Conference on Distributed and Event-based Systems, {DEBS} 2019, Darmstadt,
  Germany, June 24-28, 2019}}. \bibinfo{publisher}{{ACM}}, \bibinfo{pages}{5}.
\newblock
\showISBNx{978-1-4503-6794-3}
\urldef\tempurl%
\url{https://doi.org/10.1145/3328905.3338224}
\showDOI{\tempurl}


\bibitem[\protect\citeauthoryear{Rose}{Rose}{2012}]%
        {compressedoops}
\bibfield{author}{\bibinfo{person}{John Rose}.}
  \bibinfo{year}{2012}\natexlab{}.
\newblock \bibinfo{title}{CompressedOops}.
\newblock
  \bibinfo{howpublished}{\url{https://wiki.openjdk.java.net/display/HotSpot/CompressedOops}}.
\newblock


\bibitem[\protect\citeauthoryear{Roy}{Roy}{2007}]%
        {DBLP:journals/ipl/Roy07}
\bibfield{author}{\bibinfo{person}{B.~V: Roy}.}
  \bibinfo{year}{2007}\natexlab{}.
\newblock \showarticletitle{A short proof of optimality for the {MIN} cache
  replacement algorithm}.
\newblock \bibinfo{journal}{\emph{Inf. Process. Lett.}} \bibinfo{volume}{102},
  \bibinfo{number}{2-3} (\bibinfo{year}{2007}), \bibinfo{pages}{72--73}.
\newblock


\bibitem[\protect\citeauthoryear{Shi, Ke, Zhou, Jin, Lu, Zhang, He, Hu, and
  Wang}{Shi et~al\mbox{.}}{2019}]%
        {DBLP:journals/tocs/ShiKZJLZHHW19}
\bibfield{author}{\bibinfo{person}{Xuanhua Shi}, \bibinfo{person}{Zhixiang Ke},
  \bibinfo{person}{Yongluan Zhou}, \bibinfo{person}{Hai Jin},
  \bibinfo{person}{Lu Lu}, \bibinfo{person}{Xiong Zhang},
  \bibinfo{person}{Ligang He}, \bibinfo{person}{Zhenyu Hu}, {and}
  \bibinfo{person}{Fei Wang}.} \bibinfo{year}{2019}\natexlab{}.
\newblock \showarticletitle{Deca: {A} Garbage Collection Optimizer for
  In-Memory Data Processing}.
\newblock \bibinfo{journal}{\emph{{ACM} Trans. Comput. Syst.}}
  \bibinfo{volume}{36}, \bibinfo{number}{1} (\bibinfo{year}{2019}),
  \bibinfo{pages}{3:1--3:47}.
\newblock
\urldef\tempurl%
\url{https://doi.org/10.1145/3310361}
\showDOI{\tempurl}


\bibitem[\protect\citeauthoryear{Software}{Software}{2007}]%
        {rabbitmq}
\bibfield{author}{\bibinfo{person}{Pivotal Software}.}
  \bibinfo{year}{2007}\natexlab{}.
\newblock \bibinfo{title}{RabbitMQ}.
\newblock \bibinfo{howpublished}{\url{https://www.rabbitmq.com/}}.
\newblock


\bibitem[\protect\citeauthoryear{Suhothayan, Gajasinghe, Narangoda, Chaturanga,
  Perera, and Nanayakkara}{Suhothayan et~al\mbox{.}}{2011}]%
        {suhothayan2011siddhi}
\bibfield{author}{\bibinfo{person}{Sriskandarajah Suhothayan},
  \bibinfo{person}{Kasun Gajasinghe}, \bibinfo{person}{Isuru~Loku Narangoda},
  \bibinfo{person}{Subash Chaturanga}, \bibinfo{person}{Srinath Perera}, {and}
  \bibinfo{person}{Vishaka Nanayakkara}.} \bibinfo{year}{2011}\natexlab{}.
\newblock \showarticletitle{Siddhi: a second look at complex event processing
  architectures}. In \bibinfo{booktitle}{\emph{Proceedings of the 2011 {ACM}
  {SC} Workshop on Gateway Computing Environments, {GCE} 2011, Seattle, WA,
  USA, November 18, 2011}}, \bibfield{editor}{\bibinfo{person}{Rion Dooley},
  \bibinfo{person}{Sandro Fiore}, \bibinfo{person}{Mark~L. Green},
  \bibinfo{person}{Cameron Kiddle}, \bibinfo{person}{Suresh Marru},
  \bibinfo{person}{Marlon~E. Pierce}, \bibinfo{person}{Mary Thomas}, {and}
  \bibinfo{person}{Nancy Wilkins{-}Diehr}} (Eds.). \bibinfo{publisher}{{ACM}},
  \bibinfo{pages}{43--50}.
\newblock
\showISBNx{978-1-4503-1123-6}
\urldef\tempurl%
\url{https://doi.org/10.1145/2110486.2110493}
\showDOI{\tempurl}


\bibitem[\protect\citeauthoryear{Sybase}{Sybase}{2009}]%
        {coral8}
\bibfield{author}{\bibinfo{person}{Sybase}.} \bibinfo{year}{2009}\natexlab{}.
\newblock \bibinfo{title}{Coral8 Integration Guide}.
\newblock
  \bibinfo{howpublished}{\url{http://infocenter-archive.sybase.com/help/topic/com.sybase.infocenter.dc01030.0200/pdf/cep-IntegrationGuide.pdf}}.
\newblock


\bibitem[\protect\citeauthoryear{Theodorakis, Koliousis, Pietzuch, and
  Pirk}{Theodorakis et~al\mbox{.}}{2020}]%
        {DBLP:conf/sigmod/TheodorakisKPP20}
\bibfield{author}{\bibinfo{person}{Georgios Theodorakis},
  \bibinfo{person}{Alexandros Koliousis}, \bibinfo{person}{Peter~R. Pietzuch},
  {and} \bibinfo{person}{Holger Pirk}.} \bibinfo{year}{2020}\natexlab{}.
\newblock \showarticletitle{LightSaber: Efficient Window Aggregation on
  Multi-core Processors}. In \bibinfo{booktitle}{\emph{Proceedings of the 2020
  International Conference on Management of Data, {SIGMOD} Conference 2020,
  online conference [Portland, OR, USA], June 14-19, 2020}},
  \bibfield{editor}{\bibinfo{person}{David Maier}, \bibinfo{person}{Rachel
  Pottinger}, \bibinfo{person}{AnHai Doan}, \bibinfo{person}{Wang{-}Chiew Tan},
  \bibinfo{person}{Abdussalam Alawini}, {and} \bibinfo{person}{Hung~Q. Ngo}}
  (Eds.). \bibinfo{publisher}{{ACM}}, \bibinfo{pages}{2505--2521}.
\newblock
\showISBNx{978-1-4503-6735-6}
\urldef\tempurl%
\url{https://doi.org/10.1145/3318464.3389753}
\showDOI{\tempurl}


\bibitem[\protect\citeauthoryear{Toshniwal, Taneja, Shukla, Ramasamy, Patel,
  Kulkarni, Jackson, Gade, Fu, Donham, Bhagat, Mittal, and Ryaboy}{Toshniwal
  et~al\mbox{.}}{2014}]%
        {stormSIGMOD14}
\bibfield{author}{\bibinfo{person}{Ankit Toshniwal}, \bibinfo{person}{Siddarth
  Taneja}, \bibinfo{person}{Amit Shukla}, \bibinfo{person}{Karthikeyan
  Ramasamy}, \bibinfo{person}{Jignesh~M. Patel}, \bibinfo{person}{Sanjeev
  Kulkarni}, \bibinfo{person}{Jason Jackson}, \bibinfo{person}{Krishna Gade},
  \bibinfo{person}{Maosong Fu}, \bibinfo{person}{Jake Donham},
  \bibinfo{person}{Nikunj Bhagat}, \bibinfo{person}{Sailesh Mittal}, {and}
  \bibinfo{person}{Dmitriy~V. Ryaboy}.} \bibinfo{year}{2014}\natexlab{}.
\newblock \showarticletitle{Storm@twitter}. In
  \bibinfo{booktitle}{\emph{International Conference on Management of Data,
  {SIGMOD} 2014, Snowbird, UT, USA, June 22-27, 2014}},
  \bibfield{editor}{\bibinfo{person}{Curtis~E. Dyreson},
  \bibinfo{person}{Feifei Li}, {and} \bibinfo{person}{M.~Tamer {\"{O}}zsu}}
  (Eds.). \bibinfo{publisher}{{ACM}}, \bibinfo{pages}{147--156}.
\newblock
\urldef\tempurl%
\url{https://doi.org/10.1145/2588555.2595641}
\showDOI{\tempurl}


\bibitem[\protect\citeauthoryear{Traub, Grulich, Cuellar, Bre{\ss},
  Katsifodimos, Rabl, and Markl}{Traub et~al\mbox{.}}{2018}]%
        {DBLP:conf/icde/TraubGCBKRM18}
\bibfield{author}{\bibinfo{person}{Jonas Traub},
  \bibinfo{person}{Philipp~Marian Grulich},
  \bibinfo{person}{Alejandro~Rodriguez Cuellar}, \bibinfo{person}{Sebastian
  Bre{\ss}}, \bibinfo{person}{Asterios Katsifodimos}, \bibinfo{person}{Tilmann
  Rabl}, {and} \bibinfo{person}{Volker Markl}.}
  \bibinfo{year}{2018}\natexlab{}.
\newblock \showarticletitle{Scotty: Efficient Window Aggregation for
  Out-of-Order Stream Processing}. In \bibinfo{booktitle}{\emph{34th {IEEE}
  International Conference on Data Engineering, {ICDE} 2018, Paris, France,
  April 16-19, 2018}}. \bibinfo{publisher}{{IEEE} Computer Society},
  \bibinfo{pages}{1300--1303}.
\newblock
\showISBNx{978-1-5386-5520-7}
\urldef\tempurl%
\url{https://doi.org/10.1109/ICDE.2018.00135}
\showDOI{\tempurl}


\bibitem[\protect\citeauthoryear{Tucker, Maier, Sheard, and Fegaras}{Tucker
  et~al\mbox{.}}{2003}]%
        {watermarks}
\bibfield{author}{\bibinfo{person}{Peter~A. Tucker}, \bibinfo{person}{David
  Maier}, \bibinfo{person}{Tim Sheard}, {and} \bibinfo{person}{Leonidas
  Fegaras}.} \bibinfo{year}{2003}\natexlab{}.
\newblock \showarticletitle{Exploiting Punctuation Semantics in Continuous Data
  Streams}.
\newblock \bibinfo{journal}{\emph{{IEEE} Trans. Knowl. Data Eng.}}
  \bibinfo{volume}{15}, \bibinfo{number}{3} (\bibinfo{year}{2003}),
  \bibinfo{pages}{555--568}.
\newblock
\urldef\tempurl%
\url{https://doi.org/10.1109/TKDE.2003.1198390}
\showDOI{\tempurl}


\bibitem[\protect\citeauthoryear{Welford and Welford}{Welford and
  Welford}{1962}]%
        {Welford62noteon}
\bibfield{author}{\bibinfo{person}{Author(s) B.~P. Welford} {and}
  \bibinfo{person}{B.~P. Welford}.} \bibinfo{year}{1962}\natexlab{}.
\newblock \showarticletitle{Note on a method for calculating corrected sums of
  squares and products}.
\newblock \bibinfo{journal}{\emph{Technometrics}} (\bibinfo{year}{1962}),
  \bibinfo{pages}{419--420}.
\newblock


\bibitem[\protect\citeauthoryear{Xin and Rosen}{Xin and Rosen}{2015}]%
        {spark-memory}
\bibfield{author}{\bibinfo{person}{Reynold Xin} {and} \bibinfo{person}{Josh
  Rosen}.} \bibinfo{year}{28 April 2015}\natexlab{}.
\newblock \bibinfo{title}{Project Tungsten: Bringing Apache Spark Closer to
  Bare Metal}.
\newblock
  \bibinfo{howpublished}{\url{https://databricks.com/blog/2015/04/28/project-tungsten-bringing-spark-closer-to-bare-metal.html}}.
\newblock


\bibitem[\protect\citeauthoryear{Zagrebin}{Zagrebin}{2020}]%
        {flink-memory}
\bibfield{author}{\bibinfo{person}{Andrey Zagrebin}.} \bibinfo{year}{21 April
  2020}\natexlab{}.
\newblock \bibinfo{title}{Memory Management Improvements with Apache Flink
  1.10}.
\newblock
  \bibinfo{howpublished}{\url{https://flink.apache.org/news/2020/04/21/memory-management-improvements-flink-1.10.html}}.
\newblock


\bibitem[\protect\citeauthoryear{Zhang, Vo, Dahlmeier, and He}{Zhang
  et~al\mbox{.}}{2017}]%
        {ZhangVDH17}
\bibfield{author}{\bibinfo{person}{Shuhao Zhang}, \bibinfo{person}{Hoang~Tam
  Vo}, \bibinfo{person}{Daniel Dahlmeier}, {and} \bibinfo{person}{Bingsheng
  He}.} \bibinfo{year}{2017}\natexlab{}.
\newblock \showarticletitle{Multi-Query Optimization for Complex Event
  Processing in {SAP} {ESP}}. In \bibinfo{booktitle}{\emph{33rd {IEEE}
  International Conference on Data Engineering, {ICDE} 2017, San Diego, CA,
  USA, April 19-22, 2017}}. \bibinfo{publisher}{{IEEE} Computer Society},
  \bibinfo{pages}{1213--1224}.
\newblock
\urldef\tempurl%
\url{https://doi.org/10.1109/ICDE.2017.166}
\showDOI{\tempurl}


\end{thebibliography}

\end{document}